\documentclass[12pt,preprint]{aastex}
\shorttitle{MHD Modeling of a CME, Dimming, and a Giant Arcade}
\shortauthors{Shiota et al.}
\begin{document}
% \doublespace

\title{SELF-CONSISTENT MHD MODELING OF A CORONAL MASS EJECTION, 
CORONAL DIMMING, AND A GIANT CUSP-SHAPED ARCADE FORMATION}

\author{DAIKOU SHIOTA\altaffilmark{1}, HIROAKI ISOBE\altaffilmark{2}, 
P. F. CHEN\altaffilmark{3}, TETSUYA T. YAMAMOTO\altaffilmark{4}, \\
TAKUMA SAKAJIRI\altaffilmark{1}, AND KAZUNARI SHIBATA\altaffilmark{1}}

\altaffiltext{1}{Kwasan and Hida Observatories, Kyoto University, Yamashina, 
	Kyoto 607-8471 Japan; shiota@kwasan.kyoto-u.ac.jp}
\altaffiltext{2}{ Department of Earth and Planetary Science,
	Univerisity of Tokyo, Bunkyo-ku, Tokyo 113-0033, Japan}
\altaffiltext{3}{Department of Astronomy, 
	Nanjing University, Nanjing 210093, China}
\altaffiltext{4}{Department of Astronomy, School of Science, University 
	of Tokyo, Bunkyo-ku, 113-0033 Tokyo, Japan}

\begin{abstract}
We performed magnetohydrodynamic simulation of coronal mass 
ejections (CMEs) and associated giant arcade formations, 
and the results suggested new interpretations of observations of CMEs.
We performed two cases of the simulation: with and without heat conduction. 
Comparing between the results of the two cases, we found that  
reconnection rate in the conductive case is a little higher
than that in the adiabatic case and 
the temperature of the loop top is consistent with the theoretical value
predicted by the Yokoyama-Shibata scaling law.
The dynamical properties such as velocity and magnetic fields are 
similar in the two cases, whereas thermal properties such as 
temperature and density are very different.
In both cases, slow shocks associated with magnetic reconnection
propagate from the reconnection region 
along the magnetic field lines around the flux rope, and
the shock fronts form spiral patterns. 
Just outside the slow shocks, the plasma density decreased a great deal. 
The soft X-ray images synthesized from the numerical results are
compared with the soft X-ray images of 
a giant arcade observed with the Soft X-ray Telescope aboard {\it Yohkoh}, 
it is confirmed that the effect of heat conduction is significant
for the detailed comparison between simulation and observation.
The comparison between synthesized and observed soft X-ray images provides 
new interpretations of various features associated with CMEs 
and giant arcades. 
1) It is likely that
Y-shaped ejecting structure, observed in giant arcade 1992 January 24,
corresponds to slow and fast shocks associated with magnetic reconnection.  
2) Soft X-ray twin dimming corresponds to the rarefaction induced by 
reconnection. 
3) The inner boundary of the dimming region
corresponds to the slow shocks.
4) ``Three-part structure'' of a CME can be explained by 
our numerical results. 
5) The numerical results also suggest a backbone feature of a flare/giant 
arcade may correspond to the fast shock formed by the collision of 
the downward reconnection outflow.
\end{abstract}


\section{INTRODUCTION}

\begin{figure*}
\begin{center}
\plotone{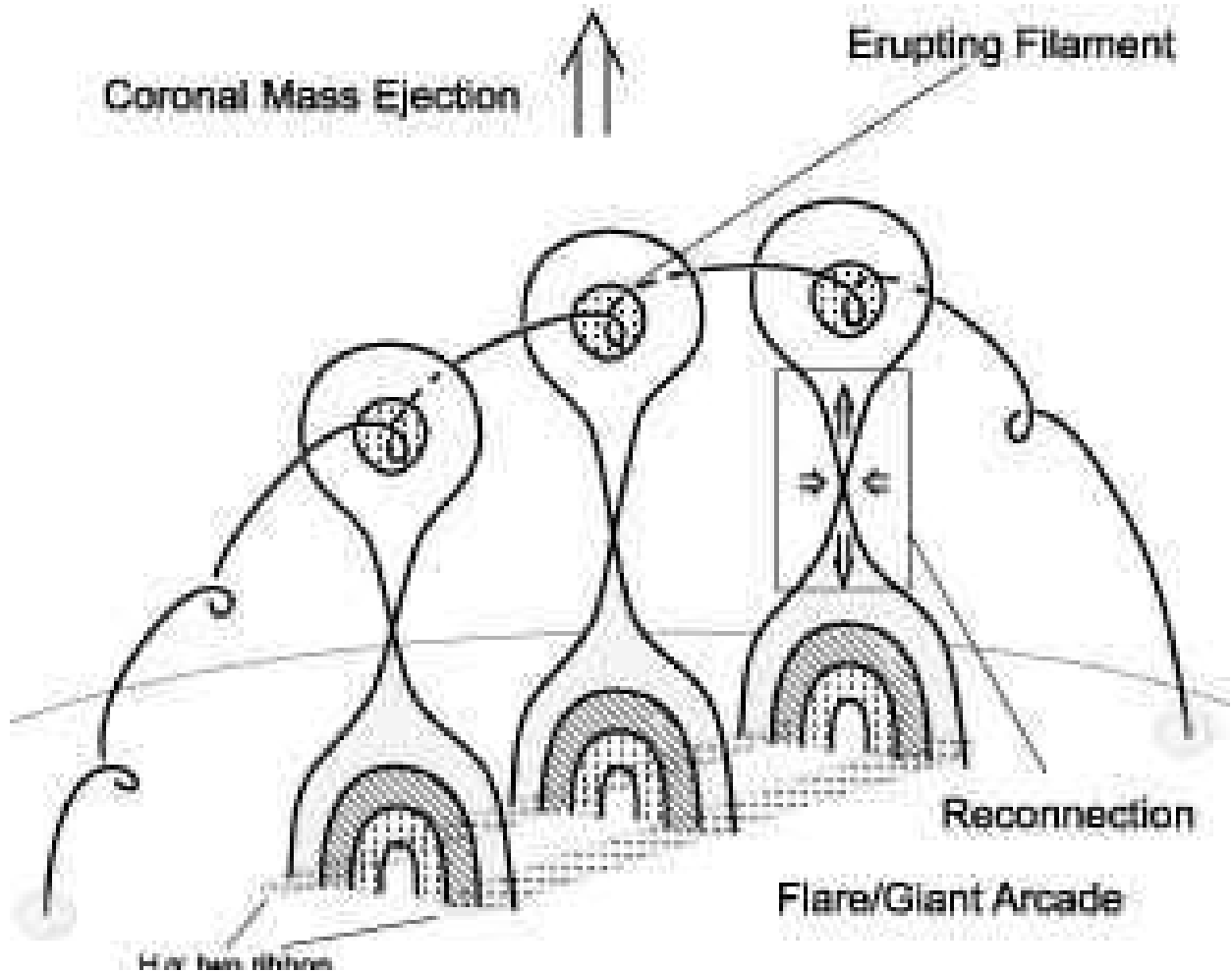}
\end{center}
\caption{Schematic picture of the standard model of the phenomenon observed 
as a CME and a flare/giant arcade.
Thick solid lines show magnetic field lines.
Note that the points where magnetic field lines cross are the X-type
neutral points where magnetic reconnection occurs.
Gray-shaded, line-shaded, and dot-shaded regions display the features 
observed in soft X-ray, extreme ultraviolet ray, and H$\alpha$, respectively. 
}
\label{flmo}
\end{figure*}

Coronal mass ejections (CMEs) are one of the most spectacular phenomena
in the solar corona.
CMEs show some kind of association with various forms of solar activities; 
40\% of CMEs were associated with flares,
but more than 70\% were associated with eruptive prominence 
or disappearing filaments (Munro et al. 1979; Webb \& Hundhausen 1987;
see Kahler 1992 for review). 
On the other hand, observations with Soft X-ray Telescope (SXT) 
aboard {\it Yohkoh} revealed that formations of giant arcades 
are associated with filament eruptions that are not associated with flares 
(Tsuneta et al. 1992; Hiei et al. 1993, 
Hanaoka et al. 1994; McAllister et al. 1996). 
Although such arcades are usually not recognized as flares because of 
their weak X-ray intensity, they have arcade structures with expanding 
cusp-shaped loops (helmet streamer), which is very 
similar to flares. From this and other evidence (e.g. Yamamoto et al. 2002)
it has been argued that their mechanism is the same as that of flares; 
magnetic reconnection. 
Considering these observations, 
Shibata (1996; 1999) suggest that 
CMEs, filament eruptions, flares/giant arcades
can be understood in a unified view:
mass ejection and magnetic energy release 
via magnetic reconnection (Fig. \ref{flmo}).

Although each CME shows different, complex structure,
many of them have a typical structure, 
so called ``three-part structure'', 
which consists of a bright core, a surrounding cavity, 
and a slightly bright outer edge (Illing \& Hundhausen 1985). 
A prominence surrounded by dark cavity is often observed
inside a helmet streamer in the pre-eruption stage, 
and this structure appears to swell once an eruption starts.
This scenario is thought to be a simple interpretation of the formation
of a three-part CME (see Hundhausen 1999 for review).
On the other hand,  Lin et al. (2004) developed 
a semi-analytical reconnection model of a CME and 
explained three-part structure formation.

On the other hand, 
{\it coronal dimmings} explained as the EUV or SXR decrease, 
are often observed around post-erupting loops 
(Sterling \& Hudson 1997; Zarro et al. 1999). 
The observation with 
Coronal Diagnostic Spectrometer (CDS) aboard 
Solar and Heliospheric Observatory (SOHO) showed 
that dimming was caused not by the decrease in temperature 
but by the decrease in plasma density 
(Harrison \& Lyons 2000).
This results indicate that the mass which existed there 
in the pre-eruption stage are lost.
Furthermore, Harra \& Sterling (2001) showed
blue-shifted motion coincided with coronal dimming 
in the CDS observations,
and therefore, suggested the lost mass is supplied to the CME. 
On the other hand, 
reconnection inflow can rarefy the inflow region
(Tsuneta 1996, Nitta at al. 2001).
It remains unclear which mechanism causes the dimmings.

In this paper we propose a new model that explains 
the formation of the three-part structure of CMEs 
and the mechanism of dimmings self-consistently.
We performed 2.5-dimensional magnetohydrodynamic 
(MHD) simulations of a coronal mass ejection 
associated with a giant cusp-shaped arcade, 
where 2.5-D means that three vector components 
are included but the space is two dimension.  
The numerical results are compared with the observations of CMEs and 
associated phenomena in order to clarify their evolution processes.

Before describing the numerical model in detail, 
it would be better to discuss the situation of the problem.
In solar flares, magnetic energy is thought to be converted into 
kinetic and thermal energies by magnetic reconnection.
In the corona, because the thermal conductivity is very high,
thermal energy is conducted very rapidly along the field lines.
When the conducted thermal energy heats the dense chromospheric plasma
(or nonthermal electrons collide to chromospheric plasma), 
the gas pressure increases suddenly.
The pressure gradient generates upward flow to the corona,
and the flare loop is filled by the plasma comes from the chromosphere.
This process is called as chromospheric evaporation (Hirayama 1974).
The flare loops filled by the hot dense plasma loses energy by radiation.

There are three key physical processes for energy balance: 
conduction, radiation, and reconnection.
The time scales of conductive cooling, 
radiative cooling, and reconnection heating are  
\begin{equation}
t_{\rm cond} \equiv \frac {3nk_{\rm B} L^2}{\kappa_0 T^{5/2}}, 
%\sim 3 \times 10^3 {\rm sec},
\end{equation}
\begin{equation}
t_{\rm rad} \equiv \frac{3nk_{\rm B} T}{n^2Q(T)},% \sim 4 \times 10^4 {\rm sec}, 
\end{equation}
and
\begin{equation}
t_{\rm rec} \equiv \frac {3 nk_{\rm B}T}{B^2 /4\pi} \frac{L}{v_{\rm A}}% \sim 97 {\rm sec},
\end{equation}
respectively, where $k_{\rm B}$ is the Boltzmann constant, 
$\kappa_0$($= 10^{-6}$ CGS) is heat conduction coefficient 
along magnetic field \citep{spit62},
$Q(T)$ is the radiative loss function (Raymond, et al. 1976). 
The time scale of reconnection heating is derived 
from Poynting flux of inflow region.
In the condition of giant arcade in quiet region (
$n = n_0 =2 \times 10^8$ cm$^{-3}$, 
$T = T_0 = 2 \times 10^6$ K, 
$L= L_0 = 2 \times 10^{10}$ cm, 
$B = 3 $ G, 
$v_{\rm A} =  B/ \sqrt{4 \pi \rho} \sim 4.7 \times 10^7$ cm s$^{-1}$,
and $Q(T) \sim 10^{-22} $ erg cm$^{3}$ s$^{-1}$; Yamamoto et al. 2002),  
$t_{\rm cond}\sim 3 \times 10^3$ sec, 
$t_{\rm rad} \sim 4 \times 10^4 $ sec, and
$t_{\rm rec} \sim 97 $ sec.
Since $t_{\rm rad}$ is much longer than $ t_{\rm rec}$, 
we can neglect the effect 
of radiative cooling for this reconnection problem. 
Though $t_{\rm cond}$ is also longer than $ t_{\rm rec}$ at the initial state,
the effect of conduction cannot be neglected 
because the temperature increases due to reconnection heating,  
so that $t_{\rm cond}$ becomes comparable to 
or even shorter than $ t_{\rm rec}$.

Although magnetic reconnection coupled with heat conduction 
is important for solar flares,
MHD simulations for such a problem 
are performed by only few researchers 
because of its numerical difficulty
(Yokoyama \& Shibata, 1997,1998, 2001, Chen et al. 1999).
These authors adopted antiparallel field for initial magnetic configuration.
For the comparison with observations, however,
we have to perform simulations in more realistic conditions.  
In our simulation, we adopted initial magnetic filed configuration 
of Chen and Shibata (2000)
and include the effect of heat conduction  
and discuss what difference the effect makes. 
The initial results of this study have already been published in 
Shiota et al. (2003).

In the next section, the numerical method is described. 
The numerical results are shown and the differences between 
adiabatic and conductive cases are described in \S 3. 
The numerical results of the conductive case are compared with 
{\it Yohkoh} observations in \S 4.
Finally, summary and discussions are given in \S 5.

\section{NUMERICAL MODELS}
\begin{figure}
\begin{center}
\plotone{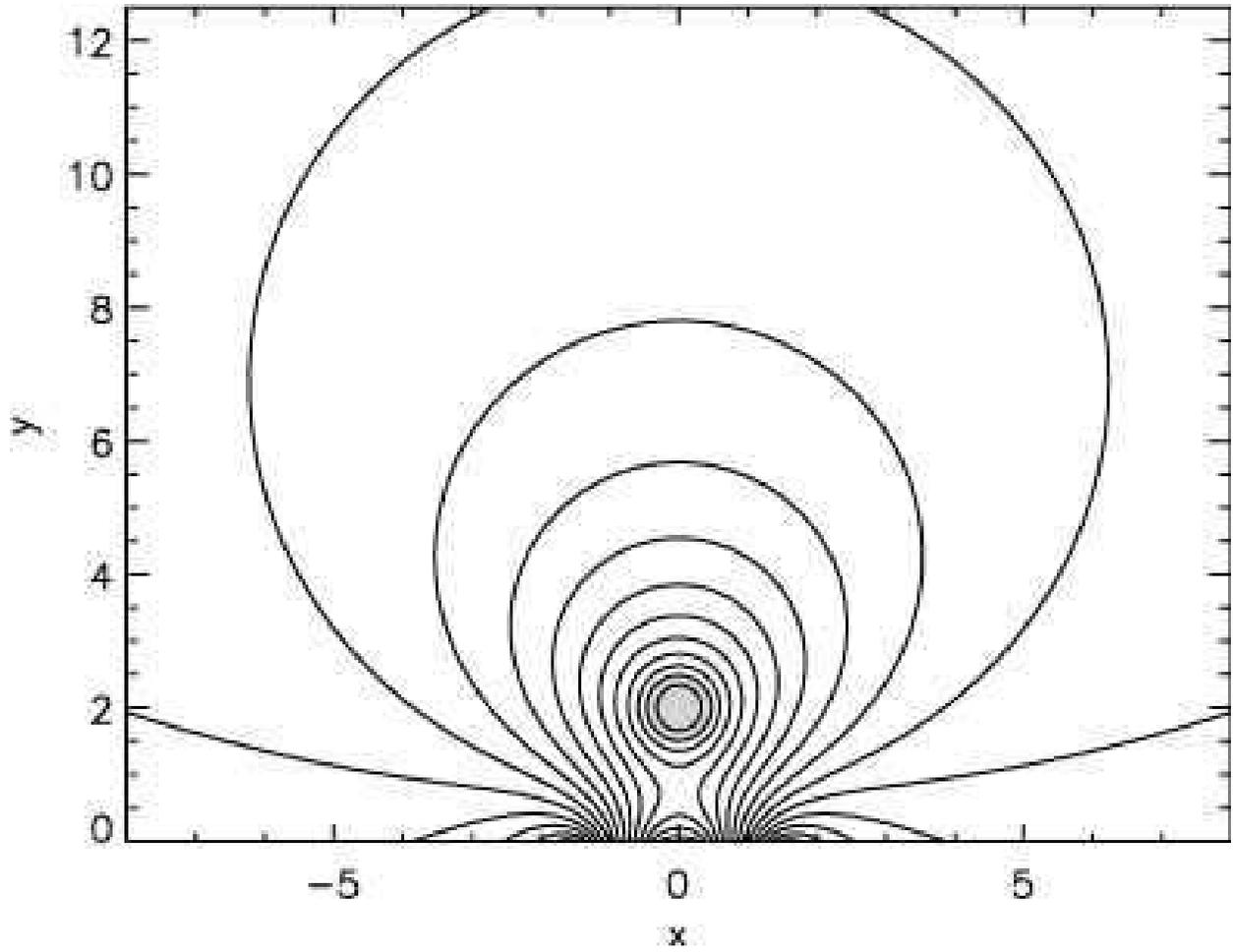}
\end{center}
\caption{Initial magnetic field configuration. Solid lines show 
magnetic field lines.
the position of the flux rope is displayed by gray shadow.
}
\label{init}
\end{figure}

\subsection{Numerical Method}

We perform simulations with a multi-step implicit scheme (Hu 1989)
for two cases: a case without heat conduction (case A) 
and a case with heat conduction (case B).
Two-and-a-half dimensional time-dependent resistive MHD equations 
considered in this study are as follows:
\begin{equation}
{\frac{\partial \rho}{\partial t}}+\nabla \cdot (\rho {\bf v})=0,
\label{eqco}
\end{equation}
\begin{equation}
{\frac{\partial {\bf v}}{\partial t}}+({\bf v}\cdot \nabla){\bf v}+
{1 \over \rho}\nabla P - \frac{2}{\beta_0} \frac{1}{\rho}{\bf j} \times {\bf B}
=0,
%-{\bf f}L_0/(\rho \rho_0v_0)=0,
\end{equation}
\begin{equation}
{\frac{\partial {\bf B}}{\partial t}}-\nabla \times ({\bf v}\times {\bf B})
+\nabla \times (\eta \nabla \times {\bf B})=0,
\end{equation}
\begin{equation}
{\frac{\partial T}{\partial t}}+{\bf v} \cdot \nabla T
+(\gamma-1)T\nabla \cdot {\bf v}-{\frac{2(\gamma-1)\eta}{\rho \beta_0}}
{\bf j}\cdot {\bf j}-C\frac{1}{\rho} Q=0,
\label{eneq}
\end{equation}
where 
${\bf B}={\nabla \times (\psi {\bf {\hat{e}}}_z)}+B_{z}{\bf {\hat{e}}}_z$,
${\bf j}=\nabla \times {\bf B}$, $\eta$ is magnetic resistivity,
the quantity $Q$ and $C$ are the dimensionless 
heat conduction function and its coefficient.
The seven independent variables are the plasma density ($\rho$), 
temperature ($T$), velocity ($v_x,v_y,v_z$), magnetic flux function ($\psi$), 
and perpendicular component of magnetic field ($B_z$).
The seven variables normalized with
the characteristic values of giant arcades are used in the equations 
(\ref{eqco}) - (\ref{eneq}).
The characteristic values of length, density, and temperature are  
obtained from the Yohkoh observations (see Yamamoto et al. 2002), 
which are $L_0 = 2 \times 10^{10}$ cm, 
$\rho_0= 3.2 \times 10^{-16}$ g cm$^{-3}$
(i.e., $n_0 =2 \times 10^8$ cm$^{-3}$),
and $ T_0 = 2 \times 10^6 $ K, respectively. 
The unit of velocity is  
\begin{equation}
v_0 = C_{\rm s} = \sqrt{\frac{2 k_{\rm B}T_0}{m_{\rm H}}}  = 
181.8 {\rm km s}^{-1},
\end{equation}
where $k_{\rm B}$ is Boltzmann constant and $m_{\rm H}$ is the mass of 
hydrogen atom.
The unit of magnetic field strength is chosen as 
\begin{equation}
 B_0=\sqrt{\frac{16\pi \rho_0 k_{\rm B} T_0}{m_{\rm H} \beta_0}} 
     = 16.66 {\rm G}.
\end{equation}  
The parameter $\beta_0$ is the ratio of gas to magnetic
pressure at the surface of the flux rope 
(magnetic configuration is descried in next subsection).    
$\beta_0$ is chosen to be 0.01.
The unit of time becomes $\tau_{\rm A0} = L_0 / v_{\rm A0} $,
where $v_{\rm A0} $ is the local Alfv\'{e}n velocity around the flux rope 
($v_{\rm A0} = B_0 /\sqrt{4 \pi \rho_0} =2571$ km s$^{-1}$) 
and $\tau_{\rm A0} = 77.8 $ s. 

In only case B, we assume anisotropic Spitzer type heat conduction,
$Q$ and $C$ in equation (\ref{eneq}) are written as follows:
\begin{equation}
Q = \left\{ \begin{array}{ll}
0, & {\rm (case A)} \\
\nabla\cdot \left({ \kappa_0 T^{5\over 2} \frac{
({\bf B}\cdot \nabla T)}{{\bf B}^2}}{\bf B}\right)
, & {\rm (case B)} \end{array} \right.
\end{equation}
\begin{equation}
       C = {\frac{2(\gamma-1)\kappa_0 T_0^{\frac{7}{2}}}{\rho_0 L_0 v_0^3}},
\end{equation}

The resistivity($\eta$) is taken as an anomalous type as follows:  
\begin{equation}
\eta=\left\{ \begin{array}{ll}
%\eta_0\min(1,|{\displaystyle \frac{v_{\rm d}}{v_{\rm c}}}|-1), 
%& |v_{\rm d}| \geq v_{\rm c},\\
\eta_0 |{\displaystyle \frac{v_{\rm d}}{v_{\rm c}}}|-1, 
& |v_{\rm d}| \geq v_{\rm c},\\
 0, & |v_{\rm d}| < v_{\rm c} ,
        \end{array}
  \right.
\label{anomal}
\end{equation}
where $v_{\rm d} \equiv {j_z}/{\rho}$ is 
the (relative ion-electron) drift velocity, 
and the dimensionless parameters are
assumed as $v_{\rm c} = 2.0$, and $\eta_0 = 0.002$.
It is known that the anomalous resistivity may be caused 
by plasma instabilities (Coppi \& Friedland 1971) in localized region,
and localized resistivity triggers fast reconnection (Ugai \& Tsuda, 1977).

The simulation box ($-8 \le x \le 8$, $0 \le y \le 12.5 $) is discretized by 
$ 321 \times 501 $ grid points, which are distributed 
non-uniformly in the $x$-direction and uniformly in the $y$-direction.
The left ($x=-8$), the right ($x=8$), and the upper boundaries
($y=12.5$) are free boundaries 
where the plasma, magnetic field, and waves could pass through freely. 
The bottom of the simulation box is a line-tying boundary
where all quantities, except for $T$, are fixed outside of 
the emerging flux region (described in section \ref{EFR}). 
$T$ is determined by extrapolation, i.e., the value was specified to be 
the same as that at its neighboring point.

\subsection{Initial Condition}

Initial magnetic field configuration is shown in Figure \ref{init}.
We adopt this configuration to mimic a cross section of   
%a ``inverse-polarity'' filament and its surroundings.
the arcade and the flux rope shown in Figure \ref{flmo}.

This configuration is presented by three separate current elements 
(see Chen \& Shibata 2000): 
a line current contained within flux rope (which determines $\psi_l$), 
its image current below the photosphere (which determines $\psi_i$), 
and four line currents just below the photosphere (which determines $\psi_b$). 
Potential field created by those currents are expressed as follows:
\begin{equation}
\psi_b= \ln{\frac {[(x+0.3)^2+(y+0.3)^2]
[(x-0.3)^2+(y+0.3)^2]}{[(x+1.5)^2+(y+0.3)^2][(x-1.5)^2+(y+0.3)^2]}},
\end{equation}
\begin{equation}
\psi_i =-{r_0 \over 2}\ln(x^2+(y+h)^2),
\end{equation}
\begin{equation}
\psi_l=\left\{ \begin{array}{ll}
r^2/(2r_0), & r \leq r_0;\\
r_0/2-r_0\ln(r_0)+r_0\ln(r), &  r > r_0.
        \end{array}
  \right.
\end{equation}
\begin{equation}
\psi=c \psi_b + \psi_l + \psi_i 
\label{psif}
\end{equation}
where  $h$ is the height of the flux rope 
and $r_0$ is the radius of the flux rope, 
which determine the configuration of magnetic field.
In this study, we set $h = 2.0$ and $r_0 =0.5$.
The coefficient $c$ in the formula (\ref{psif}) represents
the strength of background field.
This is determined by trial and error in order to guarantee that
the flux rope center approximately keeps stable for long enough time.
In this study, it is set to be 0.2534.
This physical meaning is as follows.
If we remove the flux rope, the magnetic field configuration  
are potential quadrupole field, produced by the four line currents
and the image current below the photosphere. 
This configuration makes a null point above the photosphere.
If we set at the null point a flux rope whose radius is infinitesimal,
the configuration is in unstable equilibrium.
However in the simulation, the flux rope has a finite radius
and therefore, we have to seek the solution very near to equilibrium.

To satisfy the force balance within the flux rope, a perpendicular
magnetic component (i.e., $B_z$) was introduced inside the flux rope:
\begin{equation}
B_z=\left\{ \begin{array}{ll}
 \sqrt{2(1-{r^2\over r_0^2})}, & r \leq r_0 ;\\
 0, & r > r_0,
        \end{array}
  \right.
\end{equation}
Other quantities are set to be uniform; 
$(\rho, T, v_x, v_y, v_z)=$$(1.0, 1.0, 0.0, 0.0, 0.0)$.

\subsection{CME Triggering mechanism}
\label{EFR}
Statistical studies by Feynman \& Martin (1995) show that
reconnection favorable emerging flux is correlated well with CMEs. 
The meaning of reconnection favorable is that 
the orientation of newly emerging flux is opposite to 
that of pre-existing large scale coronal magnetic field.
With numerical simulations, Chen \&  Shibata (2000) confirmed  the discovery  
of Feynman \& Martin (1995) that such emerging flux can trigger a CME.

In this study, we adopt one case of Chen-Shibata model, in which
newly emerging flux appears just below the flux rope. 
The emerging process was realized by changing the magnetic field 
at the bottom boundary within $ -0.2 \leq x \leq 0.2$.

\section{NUMERICAL RESULTS}
\begin{figure*}
%\begin{center}
\epsscale{.80}
\plotone{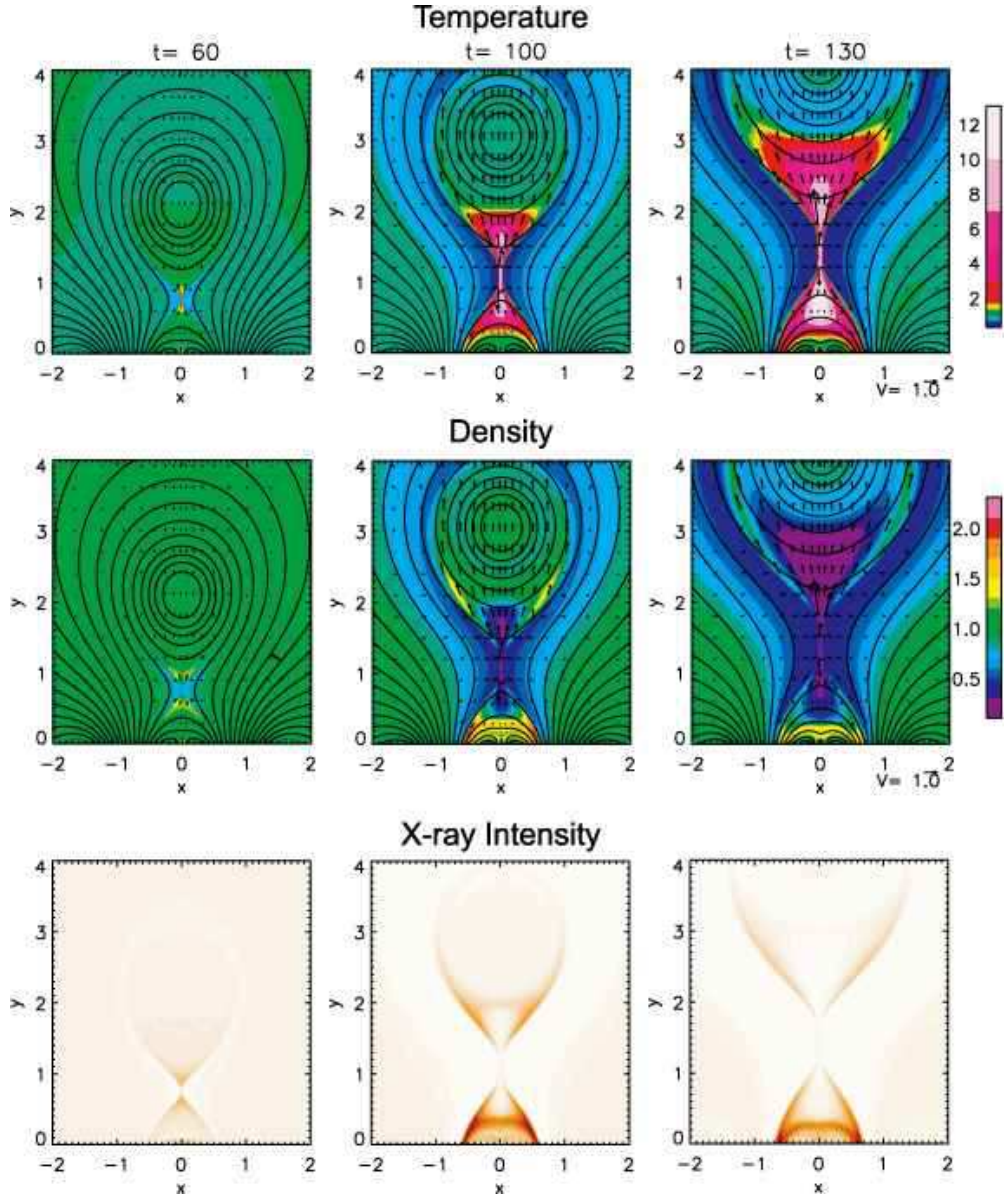}
%\end{center}
\caption{Numerical results of case A. 
Upper panels are temperature distributions at $t=60$, 100, and 130. 
Middle panels are density distributions at same times as upper ones. 
Colors display the quantities. Arrows shows velocity field at the points
and solid lines show magnetic field lines.
Lower panels are X-ray intensity distributions at same times. 
The X-ray intensity are calculated from the numerical results
after taking account of the filter response function of the Yohkoh/SXT.
The normalization units are  $\tau_{\rm A0} = 77.8 $ s, 
$L_0 = 2 \times 10^{10}$ cm, 
$n_0 =2 \times 10^8$ cm$^{-3}$,
and $ T_0 = 2 \times 10^6 $ K.
The times $t=60$, 100, and 130 mean 78, 130, and 169 
minutes from the start of the simulation.}
\label{caseA}
\end{figure*}

\begin{figure*}
\begin{center}
\epsscale{1.0}
\plotone{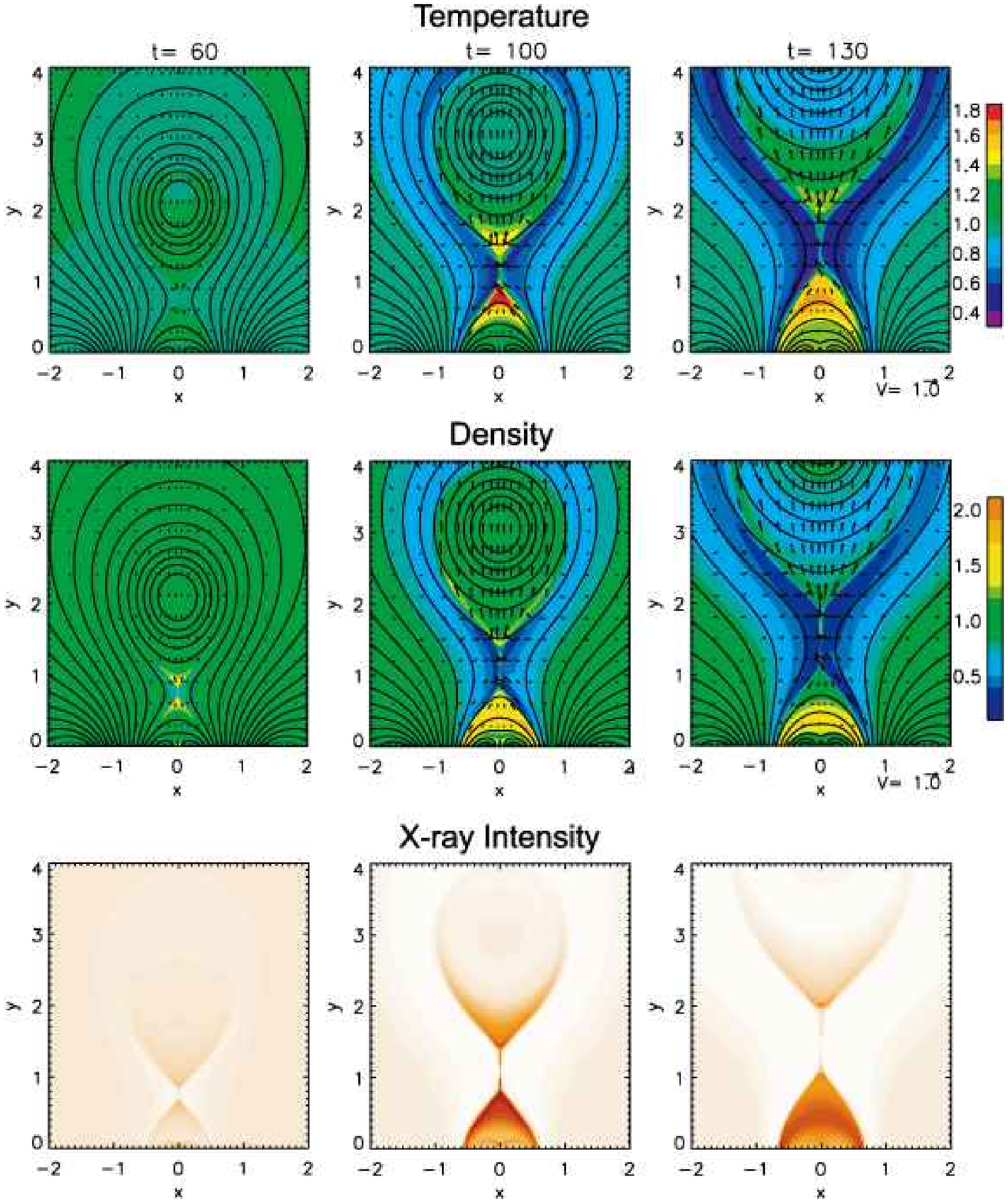}
\end{center}
\caption{Numerical results of case B. 
Temperature, density and X-ray intensity distributions are displayed
with same way as Figure \ref{caseA}.}
\label{caseB}
\end{figure*}

Time evolutions of density and temperature distributions in cases A and B 
are shown in Figures \ref{caseA} and \ref{caseB}, respectively.  

The dynamical evolutions of both cases are almost the same 
as that of case A in Chen \&  Shibata (2000).
Magnetic reconnection occurs between the newly emerged flux and 
magnetic field just above it, which leads to partial magnetic cancellation, 
and hence the decrease in magnetic pressure. 
The magnetized plasma at both sides 
(left and right to the magnetic null point [X point]) moves inward.
Then magnetic field lines at both sides are pushed in, 
and therefore a current sheet is formed near the X point.
The current density inside of the sheet increases nearly exponentially.
When the condition equation (\ref{anomal}) is satisfied,
anomalous resistivity sets in and fast reconnection occurs.
The upward outflow associated with reconnection pushes the flux rope up.
%The structure the center of which is the flux rope corresponds to a CME.

\subsection{Effect of Heat Conduction}

\begin{figure}
\begin{center}
\plotone{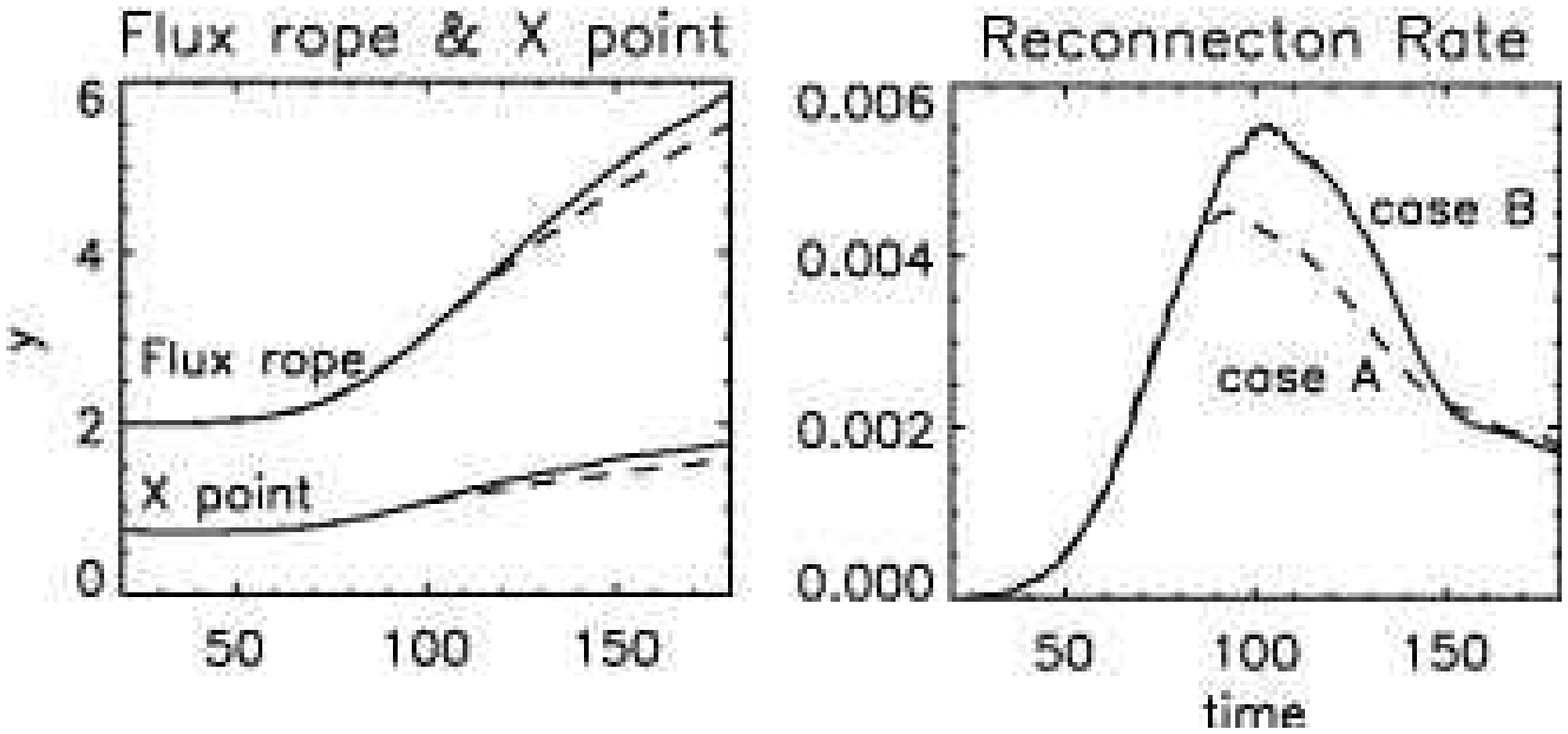}
\end{center}
\caption{Left panel shows trajectories of center of flux rope and X point 
in both cases. Right panel shows reconnection rate in both cases.
Solid lines are values in case B, dashed lines are in case A.
}
\label{rec}
\end{figure}

\begin{figure}
\begin{center}
\plotone{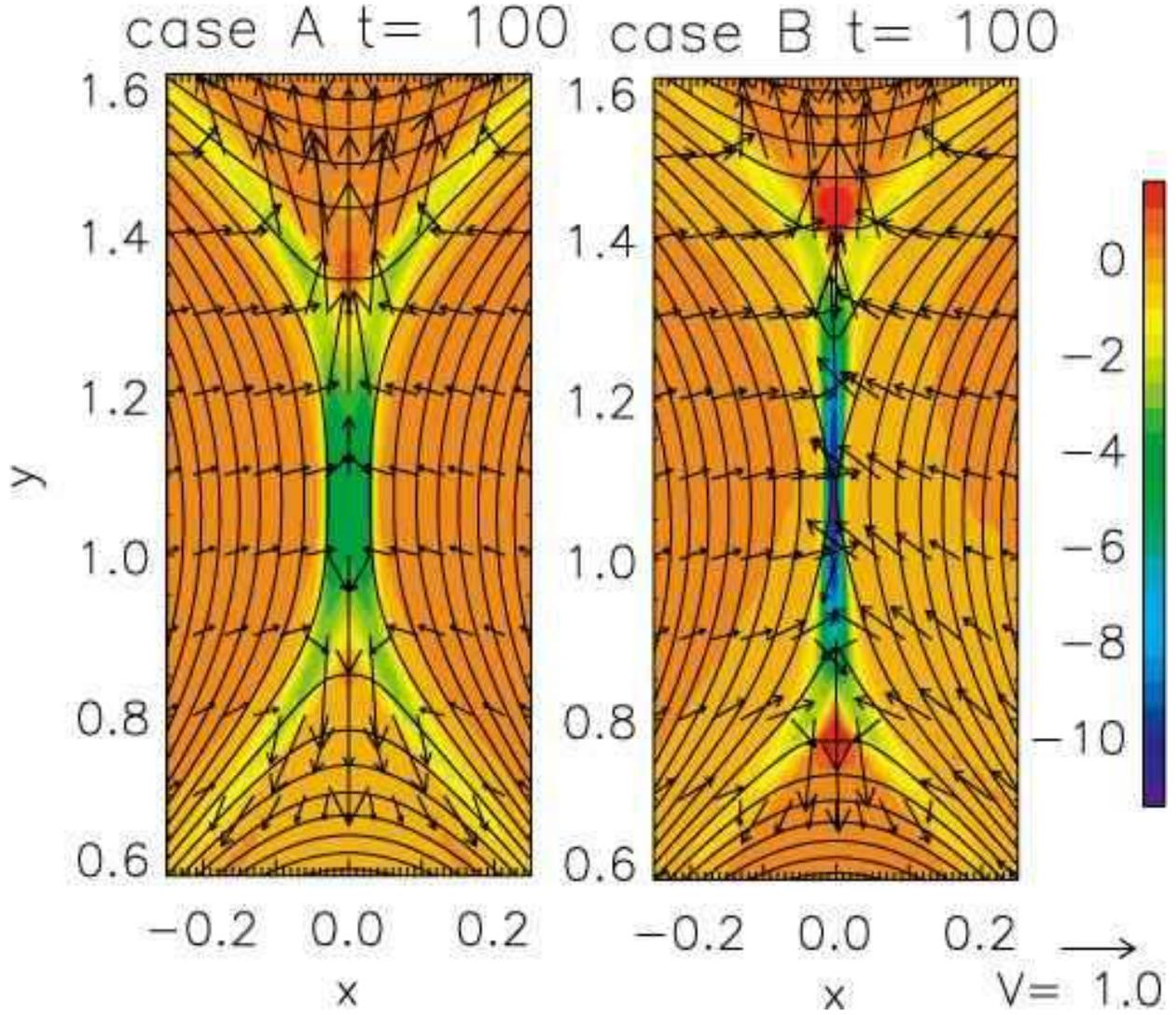}
\end{center}
\caption{Structures around the X points at $t =100$ in both cases.
Gray scales show current density distributions.  
Arrows shows velocity field at the points
and solid lines show magnetic field lines.
}
\label{Xp}
\end{figure}

\begin{figure}
%\begin{center}
\epsscale{.80}
\plotone{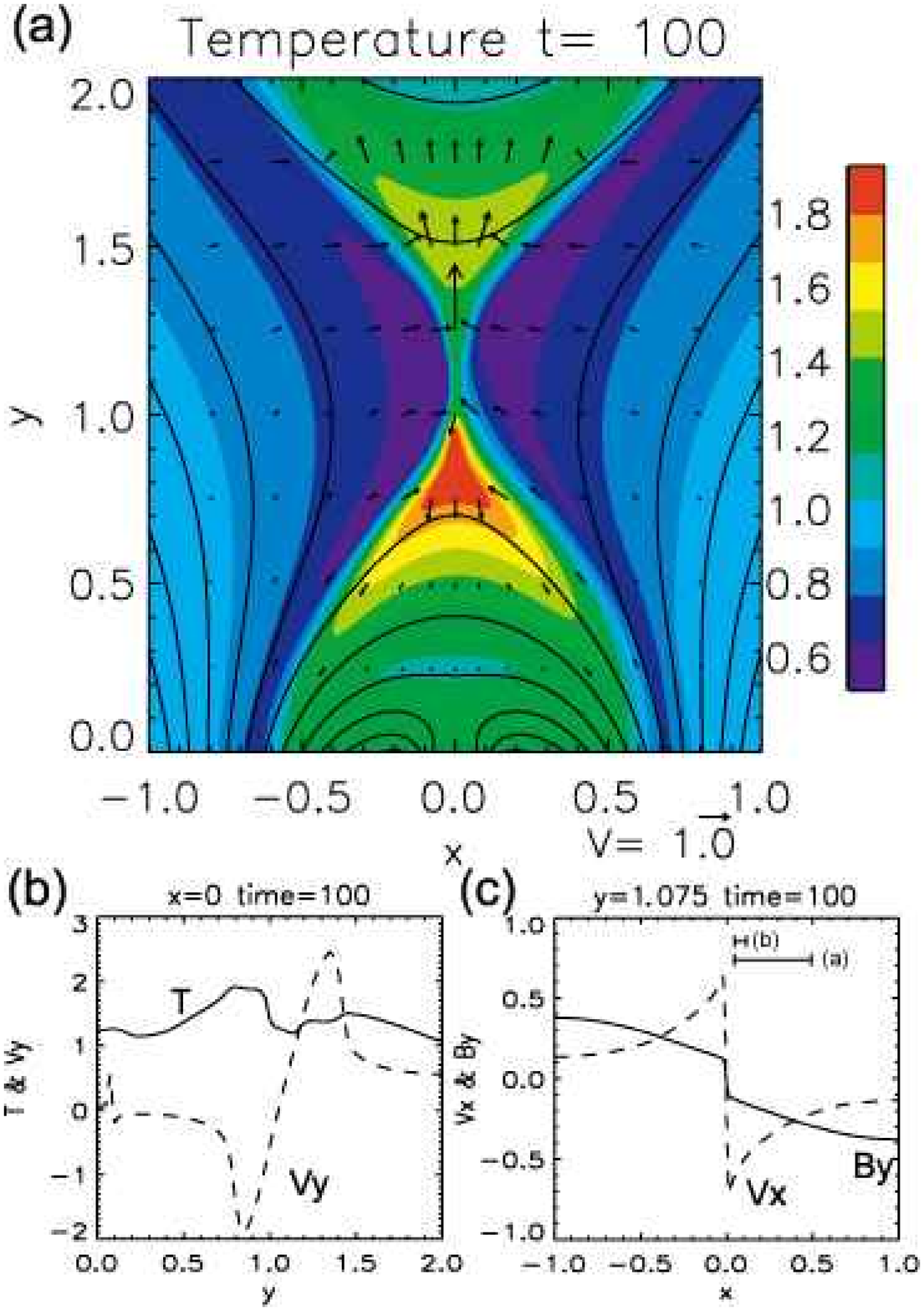}
%\end{center}
\caption{
(a) temperature distribution at $t=100$ in case B.
(b) $y$-component of velocity (dashed lines) and temperature (solid line) 
along the $y$-axis at $t=100$ in case B.
(c) $x$-component of velocity (dashed lines)  and 
$y$-component of magnetic field (solid line) 
along the line of $y=1.075$ at $t=100$ in case B.
}
\label{tscl}
\end{figure}

The clearest difference between two cases is the difference 
in density and temperature distributions within the reconnection outflows
(see Figures \ref{caseA} and \ref{caseB}). 
In both cases, the plasma that passes through the reconnection region (the 
X point) is heated very much by Joule heating.
However, in case B, because the heat conduction function become much larger in
high temperature region, the heat is conducted along the field lines.
Therefore, the temperature in reconnection region and within the reconnection 
outflows in case B becomes lower than that in case A.
The gas pressure at the reconnection region balances with  the gas and 
magnetic pressure outside the current sheet.
Magnetic field strength and gas pressure at outside of the current sheet
are almost the same in both cases.
Therefore, plasma density at reconnection region in case B become higher. 

It is found that global dynamics of the two cases are almost the same.
Figure \ref{rec}(a) shows trajectories of the flux ropes and the X points, and 
Figure \ref{rec}(b) shows reconnection rates, in both cases. 
A little difference of dynamics is caused by the difference 
in the time evolutions of reconnection rate.
Reconnection rate is defined as the reconnected magnetic flux 
per unit time and equals 
to the electric field ${\bf E} $ in the diffusion region, i.e.,
\begin{equation} 
\frac{d \psi_{\rm X point}}{dt} = {\bf E} 
\sim {\bf v}_{\rm in} \times {\bf B}_{\rm in}
\sim \eta j_{\rm X point}
\sim \eta \frac{{\bf B}_{\rm in}}{d} 
\label{eqrec}
\end{equation}
where $\psi_{\rm X point}$ and $j_{\rm X point}$
are magnetic flux function and current density at the X point, respectively, 
$ {\bf v}_{\rm in}$ and ${\bf B}_{\rm in}$ are velocity and magnetic field
of inflow region, respectively, and 
$d$ is the width of the current sheet.
The reconnection rate in case A saturated earlier than in case B.
Why reconnection rate in case A lower than in case B?
Current density distribution at $t=100$ in both case 
are displayed in Figure \ref{Xp} . 
It is found that the current density at the X point in case B is
larger than that in case A and the width of the current sheet in case B 
is thinner than that in case A. The reason is  as follows: 
In the diffusion region, plasma is heated 
by magnetic energy dissipation, and gas pressure are increased there.
The enhanced gas pressure balances the magnetic pressure 
in the inflow region. 
In case B, the heat of the inside plasma is conducted 
along the field lines, and therefore the inside pressure becomes less
than that in case A. As the sheet is pressed more and more,   
the magnetic dissipation (plasma heating) becomes larger,
and then the gas pressure balances magnetic pressure in the outside region.
In other words, the compressibility is enhanced in case B.
This is the reason why the reconnection is faster in case B. 
This results are consistent with Sato et al. (1990), 
Magara et al. (1996) and Chen et al. (1999).

Yokoyama \& Shibata (1998, 2001) performed MHD simulationeach
of magnetic reconnection coupled with heat conduction.
Considering the energy balance 
between the conduction cooling and the reconnection heating,
they derived a temperature scaling law of the top of a flare loop;
\begin{equation}
T_{\rm top} \sim  \left({\frac{B^3 L}{2 \pi \kappa_0 \sqrt{4 \pi \rho}}}\right)^{2/7}
\end{equation}
where  $B$ and $\rho$ are magnetic field strength and density 
in the inflow region, and $L$ is the half length of the loop 
which is comparable to the height of the X point.
We applied the scaling law to our numerical results.
However, in our simulation non-uniform magnetic field is assumed  
for the initial condition,
in contrast to that anti-parallel uniform magnetic field is
used in Yokoyama-Shibata's numerical simulation.
It is not easy to define a typical value of magnetic filed strength here, 
so we take the average one in the inflow region.
Because of non-uniform magnetic field,
the average magnetic field strength depends on
the area we choose as the inflow region.
We chose several cases (cases a, b, and c, as shown in Figure \ref{tscl}c)
of the definition of the inflow regions.
We tabulate the results for the definition of the inflow regions 
in Table 1. 
The values $\bar{B}$ and $\bar{\rho}$ in Table 1 are the averages
of magnetic field and density in each 
inflow region on the side of the X point 
$y=1.075$ at $t=100$ (Fig. \ref{tscl}). 
Using these values we calculate the theoretical temperature 
predicted by Yokoyama-Shibata's scaling law. 
It is found that 
the temperatures predicted by Yokoyama-Shibata's scaling law are
roughly consistent with our results.
The temperature of the loop tops may be determined by 
the condition just outside of the X point.

\clearpage

\begin{table}
\begin{center}
\caption{Temperature derived from Yokoyama-Shibata scaling law}
\label{shockv}
\begin{tabular}{lllll}
\hline
&Inflow region 	&Mag. field & Density 	&Temperature	\\ 
&& $\bar{B}$ (Gauss) & $\bar{\rho}$ (g cm$^{-3}$) &$T_{\rm YS}$ (K)\\ 
\hline
(a)&0.022$ \le x \le$0.505	 	&3.52	&1.75e-16	&1.0e+07\\
(b)&0.022$ \le x \le$0.103 		&2.25	&1.59e-16	&6.9e+06\\
(c)\tablenotemark{1} &$x=0.022$		&1.93	&1.32e-16	&6.2e+06\\
\multicolumn{4}{l}{Temperature at $(x,y)=(0, 0.75)$ in our results}	&3.6e+06\\
\hline
%\tablenotetext{1}{(c) $x=0.022$ is the just outside of the current sheet}
\end{tabular}
\tablenotetext{1}{(c) $x=0.022$ is the just outside of the current sheet}
\end{center}
\tablerefs{scal2}
\end{table}

\clearpage

\subsection{Fast and Slow MHD Shocks}
\label{sss}

\begin{figure}
%\begin{center}
%\includegraphics[width=16cm,height=18cm,keepaspectratio,clip]{fig8.eps}
\epsscale{.60}
\plotone{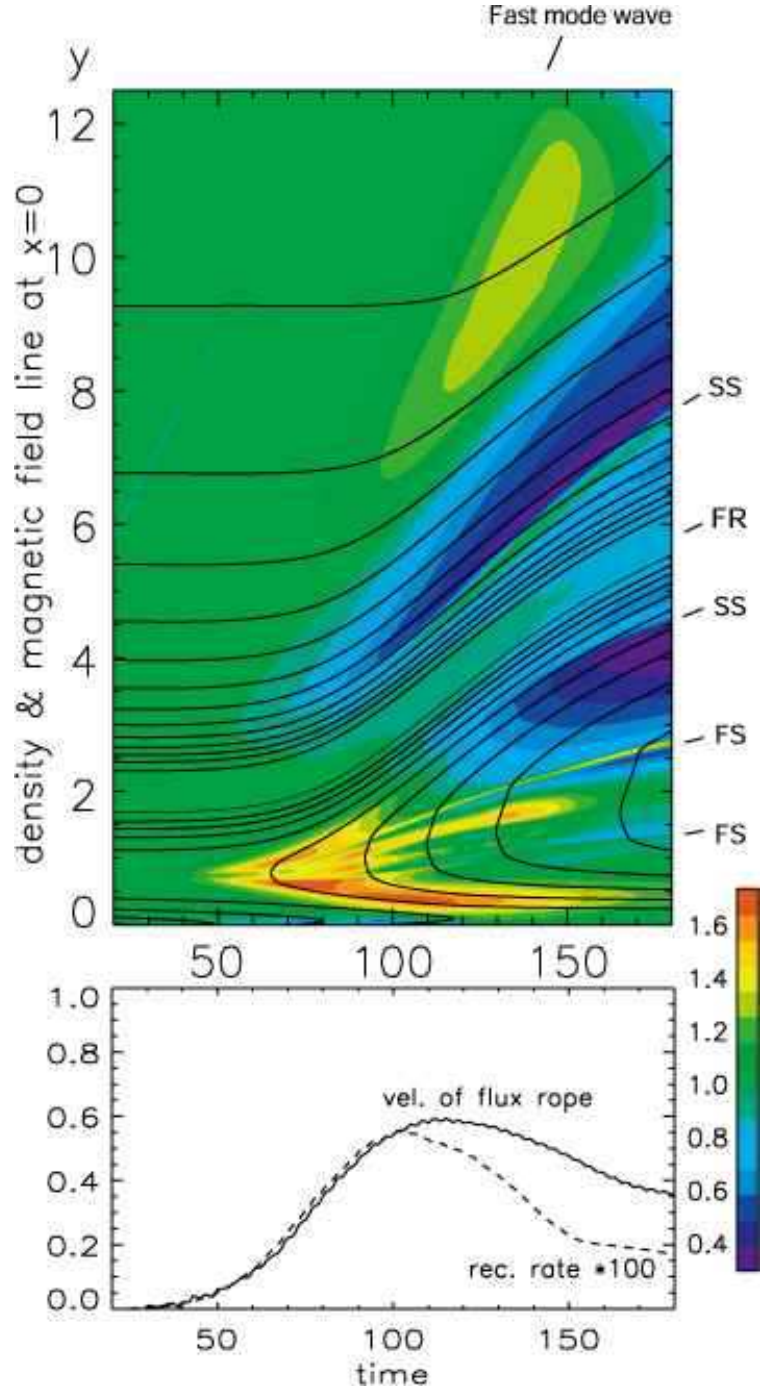}
%\end{center}
\caption{Time evolution of density at $x=0$ in case B. Upper panel displays 
time evolution of density distribution at $x=0$ with gray scales. 
Solid lines are trajectories of magnetic field lines.
Lower panel shows reconnection rate and the velocity of the flux rope 
as function of time. 
}
\label{tra}
\end{figure}

Figure \ref{tra} shows
the time evolutions of density distribution 
and the trajectory of each magnetic field line at $x=0$ (upper panel), and 
reconnection rate at the X-point and the velocity of the flux rope 
(lower panel) in case B.  
From the figures, we can find that the flux rope is accelerated
as reconnection rate increases. 
After the reconnection rate saturates, 
the flux rope is decelerated gradually.
They seem to be closely related. 
We can also find propagating density discontinuities in Figure \ref{tra},
which are also present in case A.
The highest one propagating very rapidly is the fast-mode MHD wave driven 
by the flux rope, which steepens to a fast shock.   
This correspond to the forward fast shock of a CME 
found in interplanetary observations.
The lowest two discontinuities (labeled with FS), 
through which newly appeared field lines pass, 
are formed by the collisions of reconnection outflows to 
the flux rope and the cusp-shaped loop.
They may correspond to a reverse shock just below a CME and 
a termination shock above the top of a flare loop.

\begin{figure*}
\begin{center}
\epsscale{1.0}
\plotone{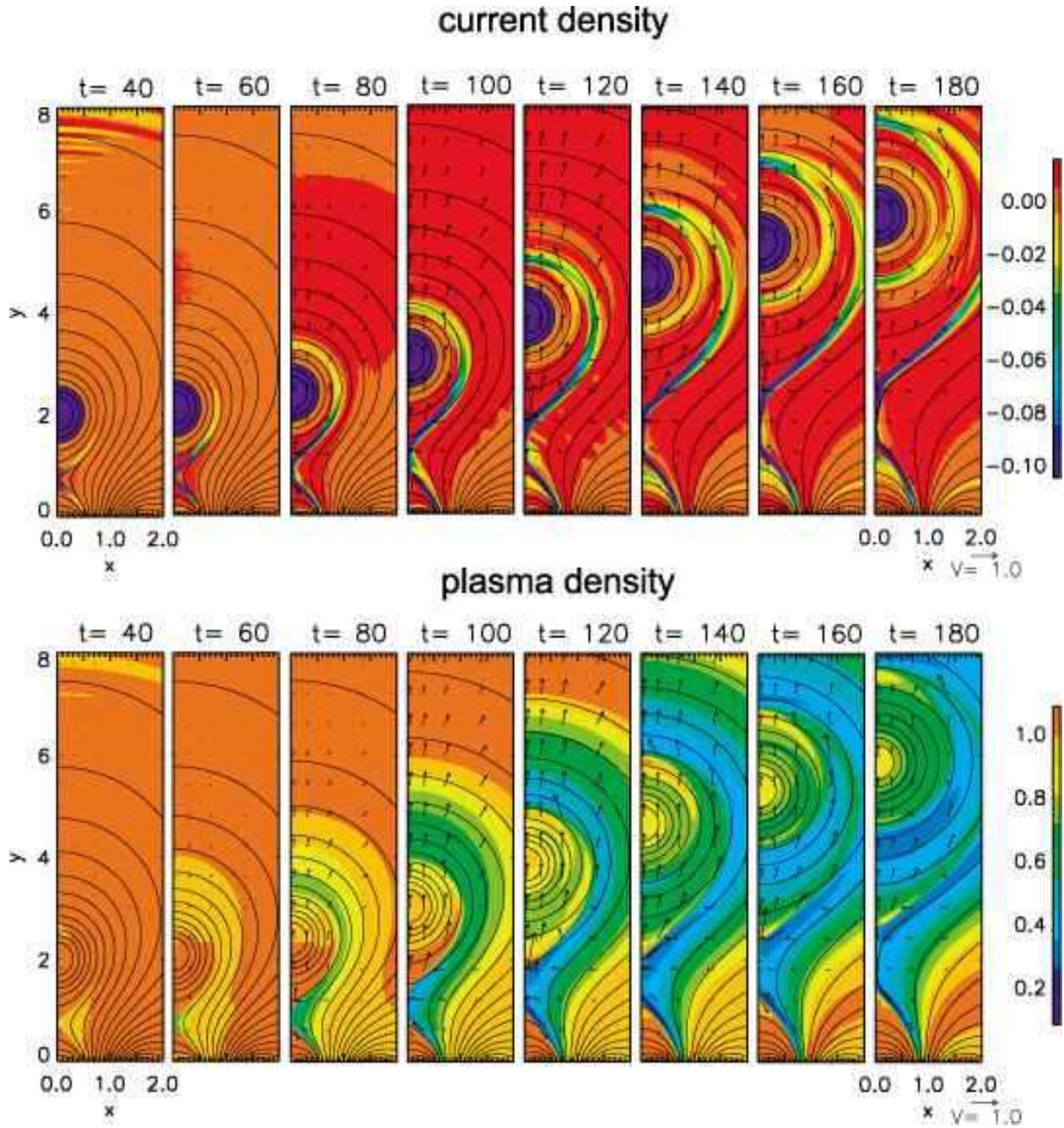}
\end{center}
\caption{Time evolutions of plasma density and current density distributions.
Colors show  plasma density and current density, respectively.
Arrows show velocity field at the points
and solid lines show magnetic field lines.
%White arrows shows positions of weak slow shocks.
}
\label{slow}
\end{figure*}

\begin{figure*}
\begin{center}
\plotone{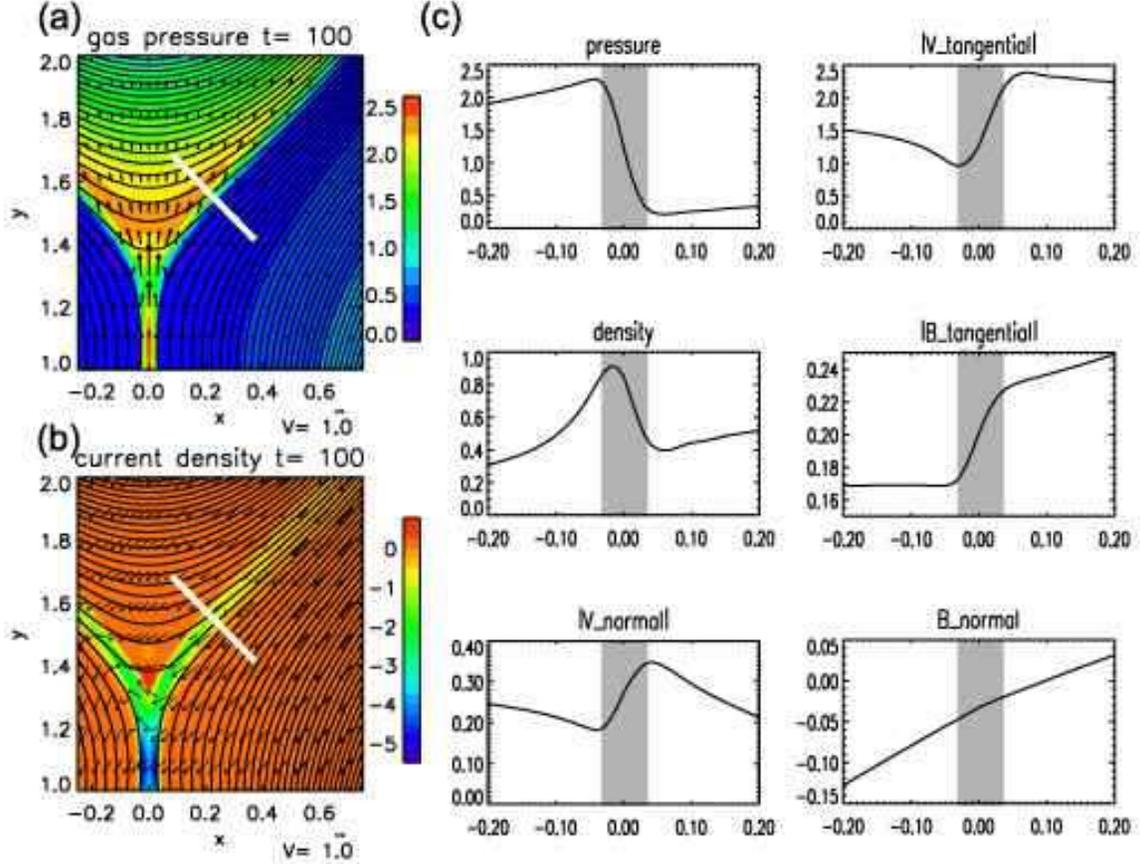}
\end{center}
\caption{Physical quantities across the discontinuity.
panels (a) and (b) are pressure and current density distributions 
at $t=100$ in case A.
Arrows show the velocity on the rest frame in panel (a)
and velocity field on the shock front rest frame in panel (b). 
Panel(c) show gas pressure, density, tangential and normal components 
of velocity and magnetic field along the white line displayed in left panels. 
}
\label{sshock}
\end{figure*}

What are the two discontinuities nearest to the center of the flux rope 
(labeled with SS) ?
Figure \ref{slow} shows the time evolutions of plasma and current densities 
distributions in case B.
Just after magnetic reconnection starts around $t=30$,  
current enhanced layers are formed along the field lines from the X point 
(separatrixes).
The layers correspond to plasma density discontinuities,
which are formed by the decrease of the outside density.
The  layers become a Y and inversed Y shaped structures,
as the flux rope rises.
Plasma velocity changes very much at these layers.
Therefore, we can identify that the layers are MHD shocks.   
In figure \ref{sshock}c, 
the distributions of several physical quantities are plotted 
(gas pressure, density, temperature, velocity, magnetic field)
along the white line shown in figures \ref{sshock}a and b.
From Rankine--Hugoniot relations for oblique MHD shocks, 
the relation between  quantities ahead (subscript with 1)
and behind (subscript with 2) of shocks
are given as follows (Priest 1982):
\begin{equation}
\frac{v_{2\perp}}{v_{1\perp}}=X^{-1}
\end{equation}
\begin{equation}
\frac{v_{2\parallel}}{v_{1\parallel}}=\frac{v_{1}^2- v_{A1}^2 }{v_{1}^2- X v_{A1}^2 } 
\end{equation}
\begin{equation}
\frac{B_{2\perp}}{B_{1\perp}}=1
\end{equation}
\begin{equation}
\frac{B_{2\parallel}}{B_{1\parallel}}=\frac{(v_{1}^2- v_{A1}^2)X }{v_{1}^2- X v_{A1}^2 } 
\end{equation}
\begin{equation}
\frac{p_{2}}{p_{1}}=X+\frac{(\gamma -1)X( v_{1}^2- v_{2}^2)}{2c_{s1}^2}
\end{equation}
where $X =\rho_2/\rho_1$,  $v_{A1}=B_1/\sqrt{4\pi \rho_1}$,
and $c_{s1}=\sqrt{\gamma p_1/\rho_1}$.
Assuming that the shadowed region (Fig. \ref{sshock}) is an MHD shock, 
we can get the values ahead of the discontinuity from the numerical results
from the theory (Table 2).
However, we cannot apply the relation exactly 
because our results present non-steady and non-uniform conditions, 
while steady and uniform condition are considered in the shock theory. 
Because nonuniform magnetic field is assumed for the initial condition,
magnetic field is stronger toward the flux rope 
and the direction of magnetic field changed gradually.
This is the reason why $B_{1\perp} < B_{2\perp}$.
Taking into account this,
we conclude that the quantities ahead and 
behind the discontinuities in the numerical results 
are roughly consistent with the shock relations and 
these discontinuities correspond to slow-mode MHD shocks 
associated with reconnection. 

\clearpage

\begin{table}
\begin{center}
\caption{Physical quantities ahead and behind of the shock}
\label{shockv_1}
\begin{tabular}{lrrr}
\hline
	& upstream	& downstream 	& downstream \\ 
	& (simulation)	&(theory)	& (simulation)	\\  \hline
$\rho$	& 0.431		& --		& 0.873		\\
$v_{\perp}$	& -0.349	& -0.172	& -0.187 	\\
$v_{\parallel}$	& -2.21	 	& -0.964 	& -0.957	\\
$B_{\perp}$	& -0.019	& -0.019	& -0.047	\\
$B_{\parallel}$	& -0.228	& -0.202	& -0.173 	\\
$p$	& 0.254		& 1.934		& 2.215 	\\  \hline
\end{tabular}
\end{center}
\end{table}

\clearpage

\begin{figure*}
\begin{center}
\plotone{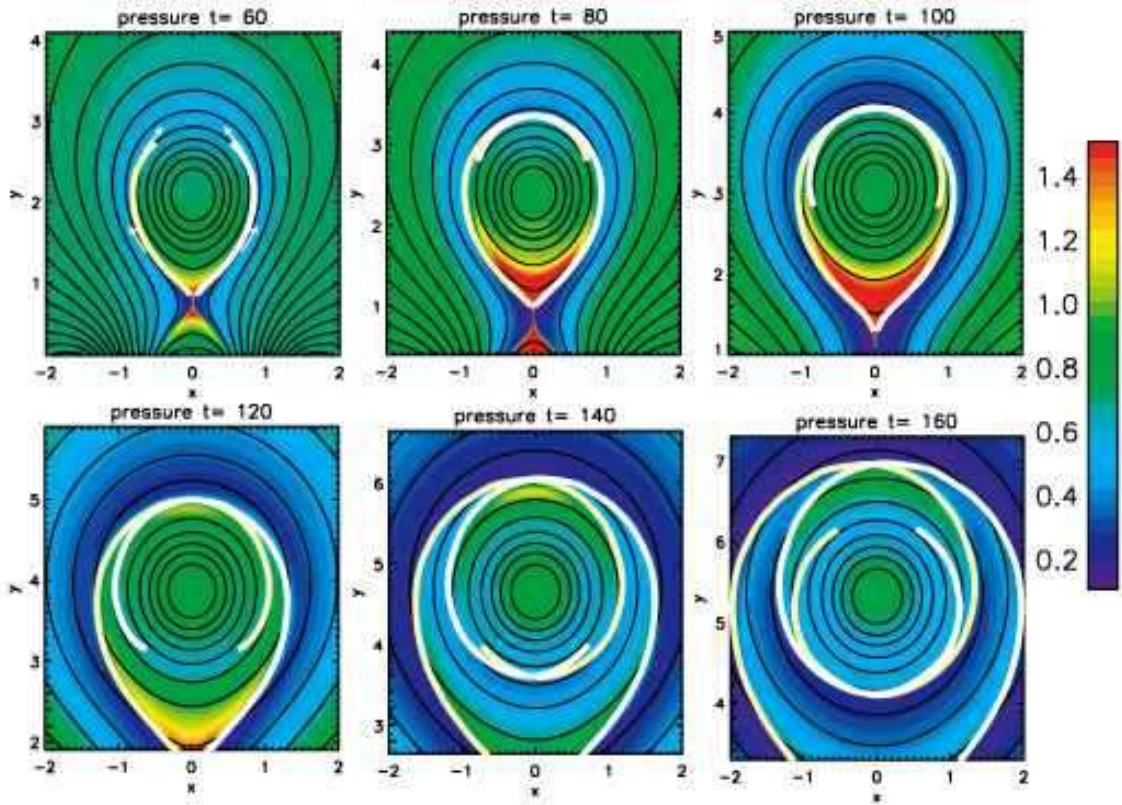}
\end{center}
\caption{The time evolution of gas pressure around the flux rope.
Colors show the gas pressure distributions. 
The centers of each panel are set to the center of the flux rope.
White and yellow curves indicate the slow shock front and white arrows 
in the panel of $t=60$ indicate the direction of the shock propagation. 
}
\label{slow2}
\end{figure*}

From Figures \ref{slow} and \ref{slow2}, 
it is found that these slow shocks continue to propagate 
around the flux rope,
the shock fronts form spiral patterns 
(white and gray lines in Fig. \ref{slow2}).
These slow shocks  correspond to  the 
two discontinuities nearest to the flux rope in Figure \ref{tra}.

\begin{figure*}
\begin{center}
\plotone{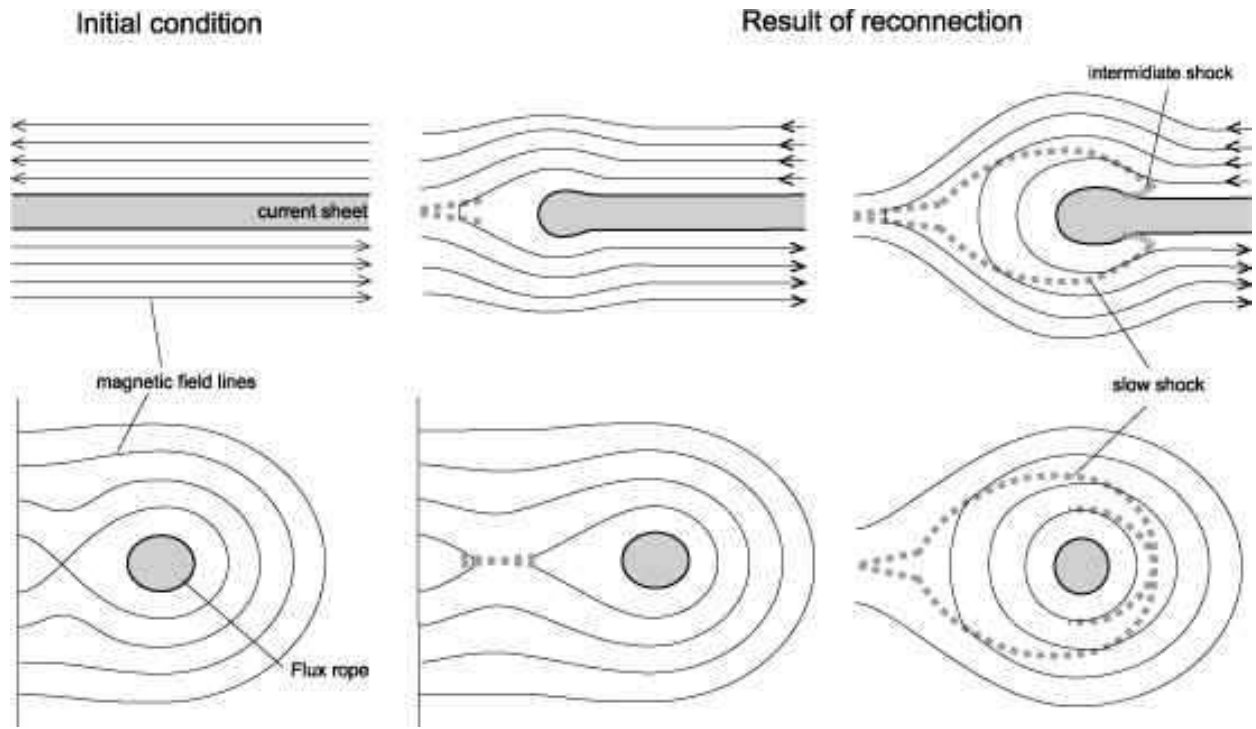}
\end{center}
\caption{
Schematic pictures of the  results of magnetic reconnection 
in an anti-parallel field model and a flux rope model.
In anti-parallel field model the slow shocks generated in 
reconnection region are terminated on intermediate shocks
propagating along the current sheet.
On the other hand, is flux rope model the current sheet does not exist,
therefore, slow shocks generated in reconnection region are not terminated. 
}
\label{slowm}
\end{figure*}

Why the slow shock structure around flux rope 
has not been found so far ?
Most of reconnection studies until now  adopted
anti-parallel magnetic field for initial condition.
Figure \ref{slowm} is a schematic pictures of the results of the simulation.
In the anti-parallel case, the results of reconnection  
forms slow shocks as shown in  upper right panels of Figure \ref{slowm}. 
The slow shocks propagate along field (along the current sheet).  
On the other hands, in our numerical simulation,
initially current sheet does not exist.
The field above the flux rope are closed.
Because the slow shock propagates along the closed field, 
the spiral patterns are formed.

\subsection{Rarefaction Structure}
\begin{figure*}
\begin{center}
\plotone{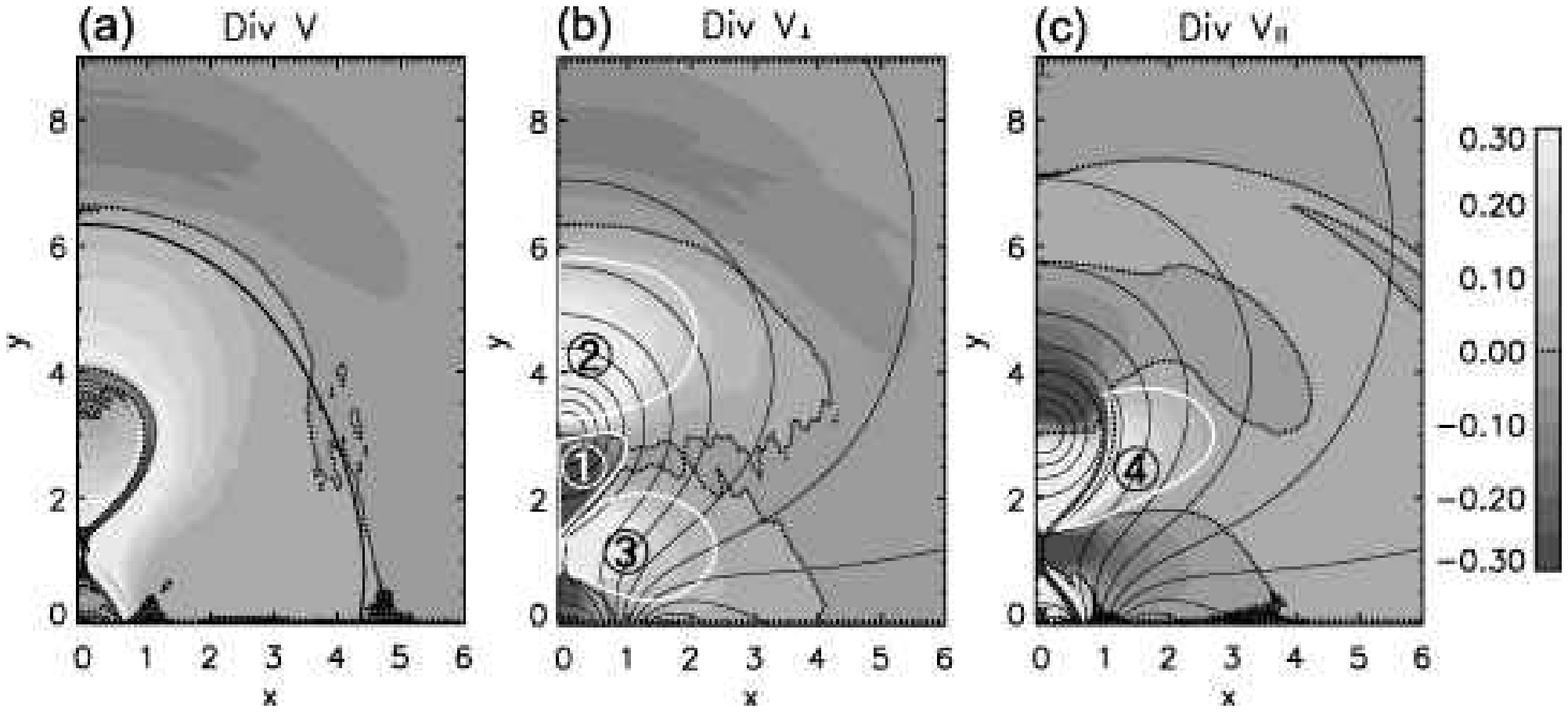}
\end{center}
\caption{ 
(a) Divergence of velocity at $t=100$, and 
thick lines show the boundaries where plasma $\beta =1$.
(b) Divergence of velocity 
perpendicular to the magnetic field line.
(c) Divergence of velocity 
parallel to the magnetic field line.
Solid lines in (b) and (c) are shown magnetic filed lines. 
Dotted lines of all panels show the boundary where the value of the divergence is equal to zero. 
}
\label{div}
\end{figure*}

We can see the region where the density decreases (rarefaction)
around the flux rope and 
the cusp-shaped loop in both cases,
just outside the slow shocks discussed above.
Why does the rarefaction occur ?

In order to examine the motion of the plasma 
we calculate the divergence of the velocity at $t=100$ (Fig. \ref{div}). 
Figure \ref{div}a shows div $v$ with gray scales where the white indicates 
divergence (expansion) and the black indicates the opposite (compression).   
In order to clarify the mass motion,	
the velocity is divided into two components
which are parallel and perpendicular to the magnetic field,
shown as follows,
\begin{equation}
{\bf v}_{\parallel} =  \frac{{\bf B} \cdot {\bf v}}{ B^2}{\bf B}
\end{equation}
\begin{equation}
{\bf v}_{\perp} =  {\bf v} - {\bf v}_{\parallel}.
\end{equation}
We also calculate the divergence of the two components of the velocity
(Fig. \ref{div} b and c). 

When fast reconnection starts, 
upward and downward directed reconnection outflows are generated. 
The upward one collides with the flux rope and compresses the magnetic field 
(region 1 displayed in Fig. \ref{div}b). 
Since this compression causes a magnetic pressure increase 
and leads to flow along the field lines. 
In fact, the region 1 shows divergence in Figure \ref{div}b.
The magnetic pressure increase in region 1 pushes the flux rope up.

The rise motion of the flux rope compresses the magnetic loops above it,
and leads to the magnetic pressure increase. 
In addition, because the magnetic field strength 
in the initial condition decreases as $ \sim y^{-3}$, 
the magnetic pressure decreases very rapidly. 
As the flux rope rises higher, 
the magnetic pressure inside the flux rope becomes larger than  
that in the ambient plasma, which leads to the expansion
to the direction perpendicular to the magnetic field
of the structure
(region 2 displayed in Fig. \ref{div}b).
This expansion is caused by the magnetic field and therefore
occurs only in the region where magnetic fields are dominant. 
In fact, the boundary between divergence and convergence is located
quite near the boundary plasma $\beta=1$ as shown in Figure \ref{div}a.  
Above the region plasma $\beta<1$, 
convergence is present because the plasma is swept up. 
%This is proposed by Forbes \& Lin (2000).

On the other hand,
fast reconnection generates reconnection inflows on 
both sides of the reconnection region.
The reconnection inflow causes the expansion of the magnetic field,
and therefore the expansion of the plasma in the perpendicular direction
(region 3 displayed in Fig. \ref{div}b).
The expansion causes a decrease of pressure, 
and hence plasma flows in along the field lines. 
In fact, region 3 shows convergence in Figure \ref{div}c.
Additionally, this converging flow causes the expansion 
in region 4 in Fig. \ref{div}c
Note that we can see a converged layer 
on the inner boundary of region 4 in Figures \ref{div}a and \ref{div}c.
This corresponds to the slow shock discussed in the previous section.

These rarefaction mechanisms work simultaneously.
The plasma, which flows into the reconnection region and the slow shocks,
flows out partly upward and the rest flows out downward. 
The mass which previously existed in the rarefied regions 
goes partly to the CME and the the rest to the cusp-shaped loop.

%\newpage
\section{Comparison between Simulations and Observations}

\begin{figure*}
\begin{center}
\plotone{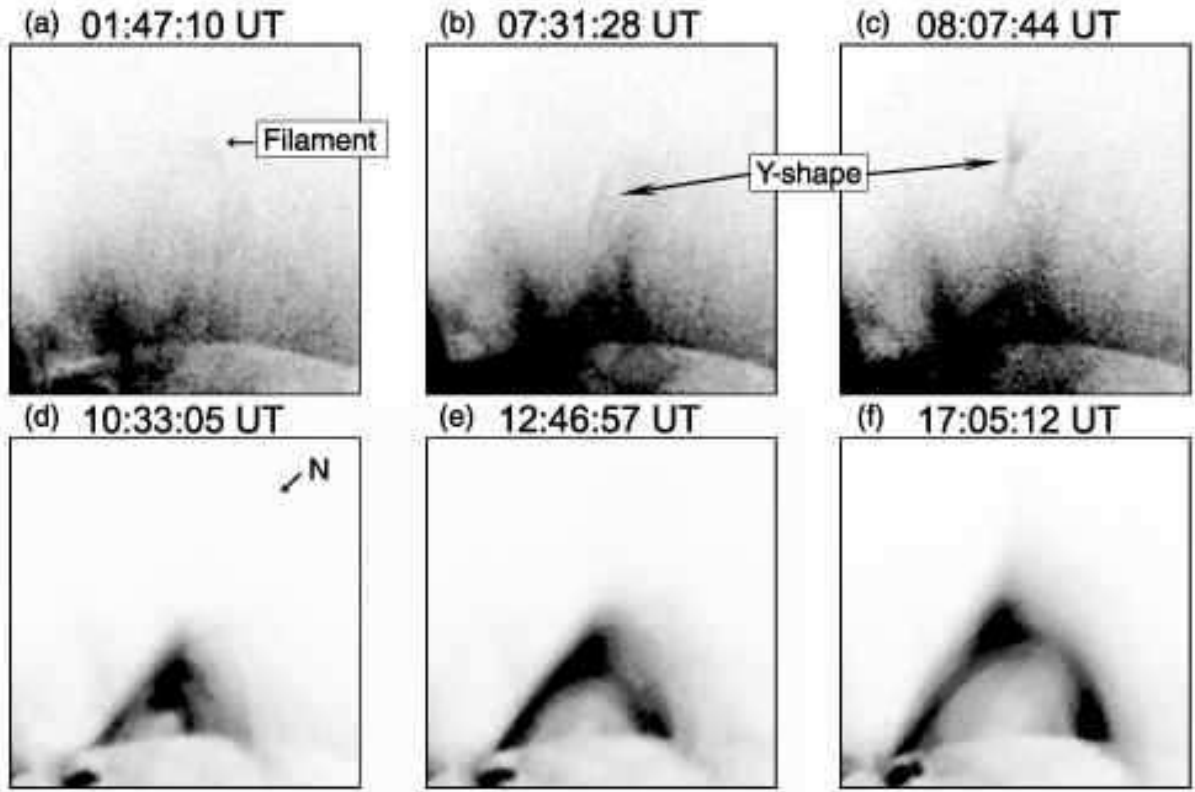}
\end{center}
\caption{Soft X-ray images of a giant arcade on 1992 January 24,  
taken with Yohkoh/SXT.
(a), (b), and (c) were taken with AlMg filter at half resolution 
($\sim 5 ^{\prime \prime}$) at 01:47:10, 07:31:28, and 08:07:44 UT.
A faint loop can be seen in panel (a), and 
it disappears before the Y-shape structure was observed.
In panels (b) and (c), the Y-shaped ejecting structure was observed. 
(d) and (e) were taken with Al.1 filter at quarter 
resolution ($\sim 10^{\prime \prime} $) at 10:33:05 UT and 
12:46:57 UT, respectively.
(f) is taken with Al.1 filter at half resolution at 17:05:11 UT.
These images are rotated with $135^{\circ}$ to counter-clockwise,
and the north direction is shown in panel (d).
The spatial scale of one side of these images are $5.4 \times 10^{10}$ cm. 
}
\label{obs}
\end{figure*}

We synthesized soft X-ray images  from the density and 
temperature in the numerical results. 
(lower panels of Figure \ref{caseA} and \ref{caseB})
These images were produced
by taking account of the filter response function of the soft X-ray telescope 
(SXT) aboard {\it Yohkoh} (Tsuneta et al. 1991).
Seen in Figure  \ref{caseA}, the density of the reconnection outflow
is very low because the temperature is very high in case A. 
Therefore, the high temperature regions in the flare loop
are observed as cavities.         
Such a flare loop have never been observed in {\it Yohkoh} observations, 
while the appearance of loop in case B is consistent with observations
(for example Fig. \ref{obs}). 
The results of adiabatic case are quite different from the real corona.
Therefore, we compared  the results of case B 
with {\it Yohkoh}/SXT  observations, in order to provide 
new interpretations of observed features.

\subsection{Y-shaped Structure}

Figure \ref{obs} shows the giant arcade (helmet streamer) formation 
on 1992 January 24 which was reported by Hiei et al. (1993).  
The event is thought to be face-on observation of 'giant arcade'
(although it has a little orientation, see Shiota et al. 2003).
In this event, after a Y-shaped structure (Fig \ref{obs}b and c) is ejected,
a cusp-shaped arch is formed, 
and then the arch seems to grow larger and larger.

These evolution in the event is consistent with our numerical results 
(Fig. \ref{caseB}).
In the numerical results, after a Y-shaped structure is ejected, 
a cusp-shaped loop seems to grow larger and larger.
%the Y-shaped structure in numerical results is formed by the slow and 
%fast shocks  associated with magnetic reconnection. 
Therefore, Shiota et al. (2003) suggest that
the Y-shaped structure in the event may correspond to the slow and 
fast shocks  associated with magnetic reconnection. 

It is known that a slow shock associated with magnetic reconnection 
 is dissociated into a conduction front and an isothermal slow shock  
in solar flares (Forbes et al. 1989; Yokoyama \& Shibata 1997). 
On the other hand, because the temperature of arcades is not as high as that 
of flares, the slow shock is dissociated slightly under the condition of 
arcades (see Shiota et al. 2003).
%Therefore,  we propose that in the arcade event, the Y-shaped structure 
%corresponds to the  slow and fast shocks associated 
%with magnetic reconnection. 

\subsection{Structure of Slow Shocks}

%(Cresmades \& Bothmer 2004, Abe \& Hoshino 2001)

\begin{figure}
\begin{center}
\plotone{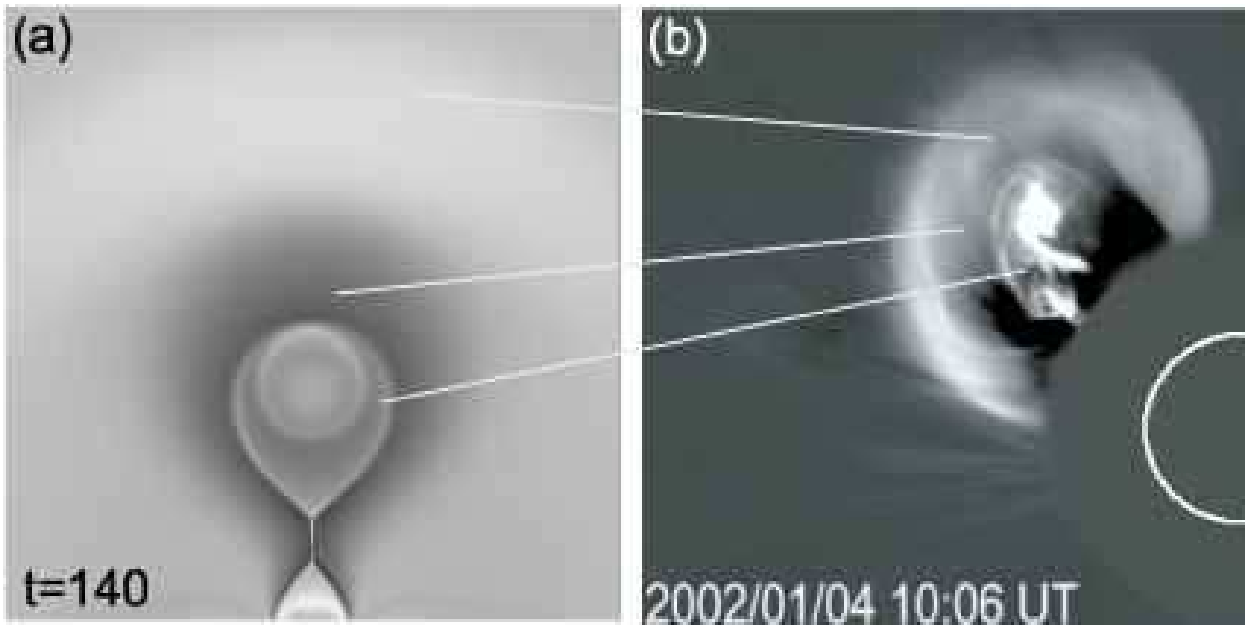}
\end{center}
\caption{
(a) density distribution at t=140 (b) LASCO C2 observation at 10:06 UT on January 4 2002 (Figure 15b of Cremades \& Bothmer 2004) 
}
\label{three}
\end{figure}

What is the outer boundary of the flux rope in the numerical results?
In Figures \ref{slow},  
the boundary is found to be a weakly current density enhanced layer, 
which corresponds to weak slow shocks (see  \S \ref{sss}).
These slow shocks propagate along the field lines 
from the X point, 
along either side of the flux rope,
and finally collide with each other above the flux rope. 
Therefore, the plasma and current densities 
are weakly enhanced just above the flux rope (Figure \ref{slow}).
After the interference, 
the shock waves also continue to propagate along the field lines. 
The flux rope continues to expand as it rises 
(see Fig. \ref{slow2} and Fig. \ref{div}a),
and the inner boundary of the expansion corresponds to
the slow shocks (Figure \ref{slow2}).
If the simulation is performed for longer time, 
the slow shocks will continue to propagate, 
producing second interference point below the flux rope.
These processes will produce shell-like structures of 
around the flux rope (Fig. \ref{three}a).   
Many CMEs show their complicated structure in SOHO/LASCO observation 
(Cremades \& Bothmer 2004, Fig. \ref{three}b).
This process may explain such  complicated 
structures of CMEs.

\subsection{Dimming}
%\subsection{Mechanism of `Dimming'}

\begin{figure*}
\begin{center}
\plotone{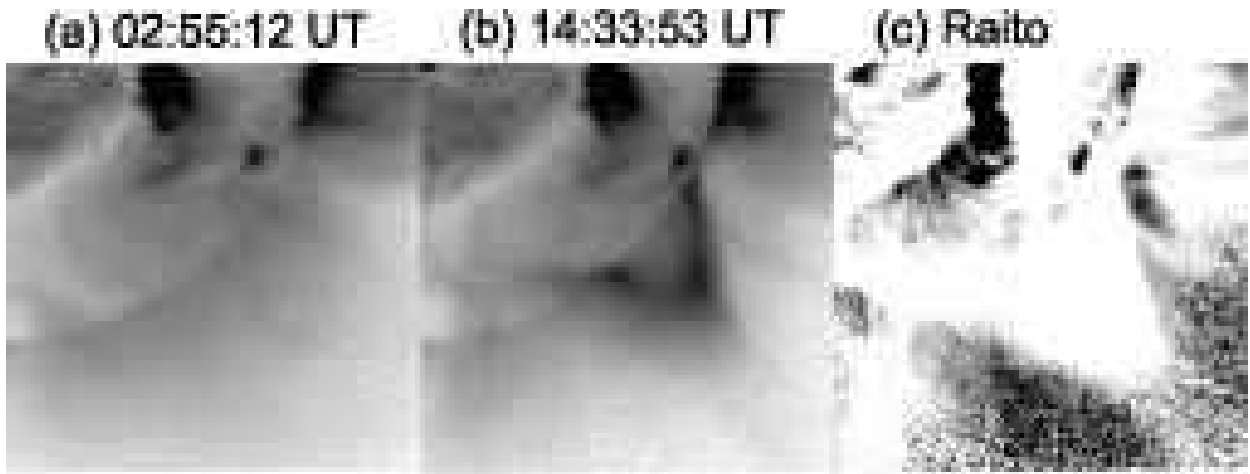}
\end{center}
\caption{Soft X-ray images (negative) taken with Yohkoh/SXT on 1992 January 24 
and the ratio of the two images (positive). 
(a) and (b) were taken with Al.1 filter at half resolution
at 02:55:12 UT and 14:33:53 UT, respectively.
(c) is the intensity ratio of (b) to (a), 
and the image displays only between zero and unity. 
}
\label{obs2}
\end{figure*}

Recent observations reveal that solar flares often occur associated with 
the decrease of soft X-ray and EUV intensity,
that is so-called {\it dimming}
(Sterling \& Hudson 1997; Zarro et al. 1999).
Hudson (1996) showed that, 
also in the giant arcade on 1992 January 24, 
dimming occurred in the regions above the cusp-shaped arcades
(shown in Fig. \ref{obs2}).
SOHO observations indicated that the 
dimming is caused by the decrease in 
the plasma density not by the decrease in temperature 
(Harrison \& Lyons 2000).
Furthermore, Harra \& Sterling (2001) showed
blue-shifted motion coincided with coronal dimming 
in the CDS observations,
and therefore, suggested the lost mass is supplied to the CME,
while the mechanism has not been confirmed.
Especially, the strong dimming is often observed 
in both sides of flare/giant arcades (Sterling \& Hudson 1997; Hudson 1996), 
hereafter we call such strong dimming as 'twin dimming'

We can see that the density around the flux rope and 
the cusp-shaped loop decreases in the simulation.
The decrease in density is observed as soft X-ray dimming (Fig. \ref{caseB}).
The dimming can be seen in both side of the X-ray arcade in the simulation, 
and these positions are consistent with those of twin dimming in observations. 
Therefore,  soft X-ray twin dimming may correspond to the rarefaction
in the simulation.
As discussed in section 3.3, the rarefaction caused by 
reconnection inflow and the expansion of the flux rope.
However, the expansion occurred mainly above the flux rope. 
Therefore the rarefaction in both side of the loop (i.e. twin dimming) 
are mainly caused by reconnection inflow. 

\subsection{Three-part Structure}

\begin{figure*}
\begin{center}
\plotone{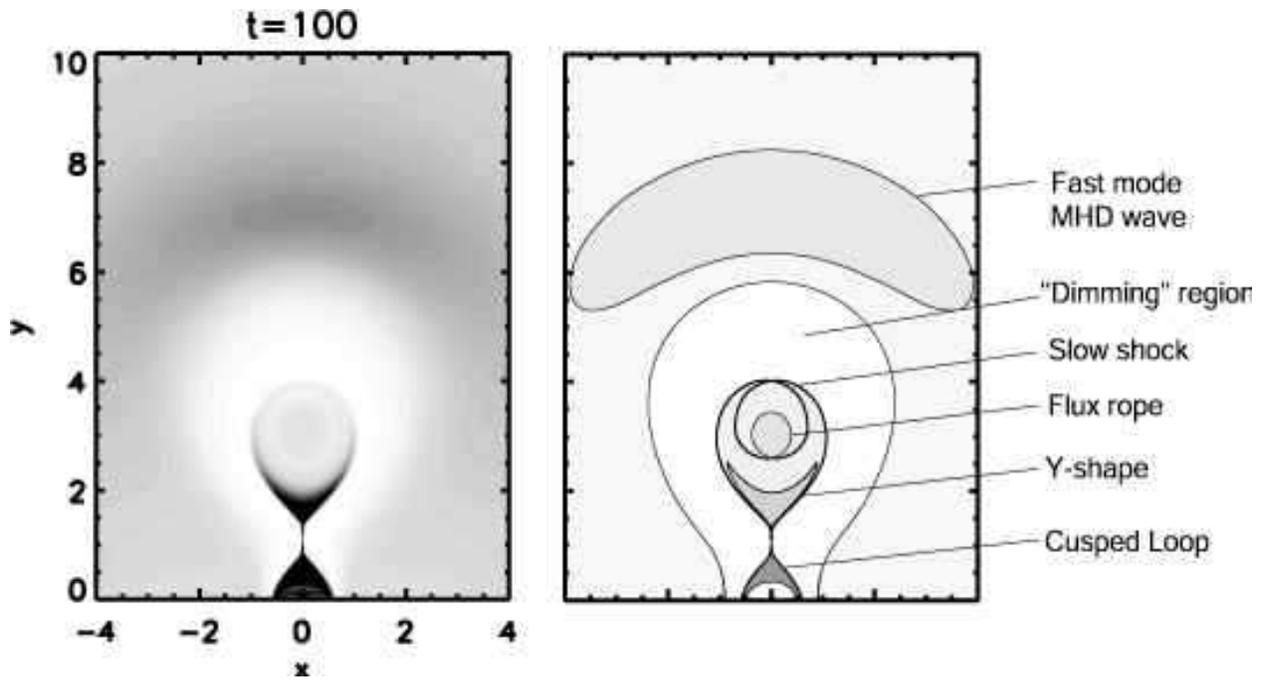}
\end{center}
\caption{Simulated negative x-ray images. 
Left panel shows calculated X-ray image at $t=100$ in case B.
Right panels show simplified same image in which each features
are indicated.
}
\label{mod}
\end{figure*}

From the point of view of CME structure, the numerical results provide 
some new interpretations. Figure \ref{mod} shows global structure
seen in the results, which is very consistent with the 
``three part structure'' observed in many CMEs.
According to the numerical results,
the core consists of the flux rope and slow shocks,
the cavity corresponds the `dimming' region around the flux rope
which connected to the twin dimming regions, 
and the front loop corresponds to the region compressed 
by the forward fast shock (Fig. \ref{mod}).
In this results, the loop in front of the CME is very weak and not a shock, 
because density is not stratified owing to the absence of the gravity.
In another case of our numerical result including gravity 
(see Shiota et al. 2004),
the wave grows to a shock easily. 

Lin et al. (2004) modeled a CME with semi-analytic reconnection model and 
explained the ``three part structure'' of CME as a separatrix bubble.
They discussed, in the paper, that the leading edge of three-part structure
corresponds to the separatrix connected to the reconnection region.
As they mentioned in the paper,  
their model is not realistic therefore more detailed computation is required.  
Our results can be applied the requirements,
and suggest new interpretations (Fig. \ref{three} and  \ref{mod})
successfully.

\subsection{Backbone Feature of Flare Arcade}

\begin{figure*}
\begin{center}
\plotone{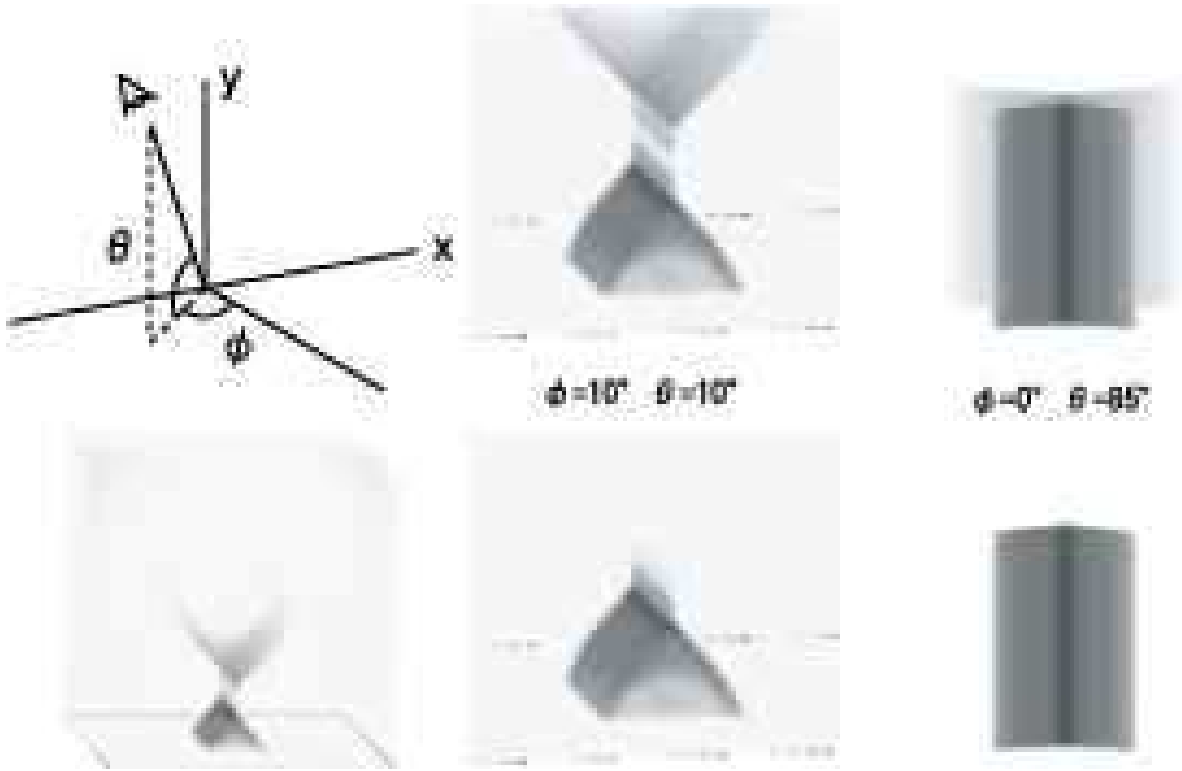}
\end{center}
\caption{
Three dimensional views of a numerical arcade at $t=100$.
The arcade is calculated with an assumption that the numerical results
are uniformly arranged to the perpendicular direction.
Upper panels are  synthesized using numerical results with out any process. 
Lower panels are synthesized using data in which we remove Y-shaped ejection 
because such Y-shaped ejection cannot be observed in most arcade flare. 
Left panels are synthesized as viewed with a horizontal angle ($\phi$) of $10^{\circ}$
and with inclination angle of angle ($\theta$) of $10^{\circ}$.
Right panels are synthesized with a horizontal angle ($\phi$) of $0^{\circ}$ 
and with inclination angle of angle  ($\theta$) of $85^{\circ}$.
}
\label{3da}
\end{figure*}

In some large arcade flares (and some giant arcades), 
we can see often backbone-like features,
which may be a series of brighter features on the top of each loop
in the arcade in soft X-ray images (Sakurai et al. 1992)  
while loop top bright features at the loop top
are also observed in some compact flares, 
which are called as `Feldman blob' (see Acton et al. 1992).
In the event on 1992 January 24, the top of the cusp-shaped loop
is brighter than other region of the loop (Figure \ref{obs}d, e, and f).
On the other hand, calculated X-ray images show brighter feature on
the top of the cusp-shaped loops. 
The brighter feature is formed by the compression in the fast shock
formed by the collision of the downward reconnection outflow.

In order to compare the numerical result with SXT observations more directly,
we calculated 3-dimensional views of the numerical results, 
assuming that the numerical results distributed uniformly 
to the perpendicular direction.
We apply the same way to calculate a 3-dimensional views 
as used in  Forbes  \& Acton (1996).
The calculated images are shown in Figure \ref{3da}.
However, the effect of gravity (density stratification) 
is not included in our numerical simulations.
If the effect of gravity is included, 
the density in upper corona becomes much smaller than thiese results.
The soft X-ray intensity is proportional to $n^2$, 
the brightness of the feature in upper corona becomes much smaller.
In fact, the center of the Y-shape is located at twice of the 
scale height, the intensity of the Y-shape is negligible.
Therefore, we synthesized 3-dimensional soft X-ray images, 
in which the Y-shape is removed 
(the lower right two panels of Fig. \ref{3da}). 
In the figure,  the bright feature in the numerical results
looks like a backbone feature.
The results, hence,  indicate that a backbone like feature of flare/giant 
arcades may correspond to the plasma compressed by the fast shock
produced by a reconnection outflow.

\section{CONCLUSION}

In this paper, we proposed a new model that explains 
the formation of the three-part structure of CMEs 
and the mechanism of dimmings self-consistently.

we performed MHD simulations 
of CMEs associated with giant cusp-shaped arcades for the two cases;
one case includes heat conduction effect, 
while the other does not include. 
The dynamics is similar in both case,
while the reconnection rate in the conductive case is 
enhanced slightly compared to the adiabatic case 
and the temperature of the loop top is consistent with the theoretical value
predicted by the Yokoyama-Shibata scaling law.
In the conductive case, a slow shock
is hardly dissociated to an isothermal 
slow shock and a conduction front,
different from the case for solar flares.
As magnetic reconnection occurs slow and fast shocks are formed 
to be attached to  a Y-shaped structure and a cusp-shaped loop.
The slow shocks continue to propagate
along the field lines around the flux rope, forming two spiral patterns.   
The density just outside the slow shocks decreases a great deal.
This decrease in density (rarefaction) is caused 
by expansion and reconnection inflow simultaneously. 

Compared with {\it Yohkoh}/SXT observations, 
the numerical results in case B
are consistent with the observation of a giant arcade on 1992 January 24.
A Y-shaped erupting structure observed in the event may correspond to
the slow and fast shocks associated with the magnetic reconnection
(this is reported by Shiota et al. 2003). 
Soft X-ray dimming is also observed in this event.
The position of the dimming regions are consistent with 
that of the rarefaction in the numerical results.
Since the rarefaction occurs in the region outside the slow shocks, 
the inner boundary between the ``dimming'' and ``non-dimming'' regions
may be slow shocks associated with the reconnection.

The global structure in the numerical results is similar to  the 
``three part structure'' of CMEs.
In the numerical results, the leading edge, cavity, and core correspond to 
the region compressed by piston-driven fast MHD wave,
the rarefaction region, and the flux rope.   
The slow shocks associated with magnetic reconnection 
may play a role forming in complex features around the core.

According to observations
the top of cusp-shaped loops in an arcade is brighter
than other regions, which looks like a backbone of the arcade.
Such a feature is reproduced in the numerical simulation, 
which indicates that the backbone feature corresponds to
the fast shock created by the collision between 
the reconnection outflow and the arcade.

The new results of this work are as follows.

1. We found the evolution of slow shocks associated with reconnection 
in the configuration initially a flux rope exits.
The numerical results indicate that complex structures around a CME core 
may be formed by these slow shocks. 

2. We proposed a self-consistent model of `dimming'.  
The decrease in density observed as `dimming' is caused by 
expansion and reconnection inflow simultaneously.
In the numerical results,
the cavity of the three-part CME is connected to the `dimming' regions,
and therefore the density decrease mechanisms are the same. 
We proposed a new interpretation of three-part structure of CMEs 
(Fig. \ref{mod}). 

3. The backbone-like features observed in many flare/giant arcades 
can be explained as the fast shock created by the collision between 
the reconnection outflow and the arcade.

%\bigskip

%Although our numerical results give 
%many new interpretations of observations,
%this model contains some remaining problems.
%(1) This model is two and a half dimensional model.
%Assuming to be uniform to the direction perpendicular to the plane
%(2) Gravity should be included and spherical coordinate system should
%be applied for comparison with SOHO/LASCO. 
%(3) Chromospheric evaporation should be included in order to be compared
%with observations of flares.

\acknowledgments

The authors thank D. H. Brooks, J. Lin, T. Miyagoshi, A. Asai, 
H. Tonooka, and T. Yokoyama, for useful discussions. 
The Yohkoh satellite is a Japanese national project, launched
and operated by ISAS, involving many domestic institutions,
and with international collaboration of US and  UK.
D. S. is supported by Research Fellowship of 
the Japan Society for the Promotion of Science (JSPS) for Young Scientist. 
The visit of P. F. C. to Kyoto University was supported by the JSPS.
P. F. C. is supported by NSFC (10333040 and 10403003).
This work is supported by the Grant-in-Aid for the 21st Century COE "Center 
for Diversity and Universality in Physics" from the Ministry of Education, 
Culture, Sports, Science and Technology (MEXT) of Japan.
This work was partly supported by the Grants-in Aid of the MEXT of 
Japan (16340058, KS).
This work was supported in part by the JSPS Japan-UK Cooperation Science 
Program (principal investigators: K. Shibata and N. O. Weiss).

%\appendix


\begin{thebibliography}{}
%\bibitem[Abe, Hoshino(2001)]{abe01}
%	Abe, S. A. \& Hoshino, M. 2001, Earth, Planet, and Space, 53, 663
\bibitem[Acton et al.(1992)]{acton92}
	 Acton et al. 1992, 1992, \pasj, 44, L71
\bibitem[Chen et al.(1999)]{chen99} 
	Chen, P. F., Fang, C., Ding, M. D., \& Tang, Y. H. 1999, \apj, 520, 853
\bibitem[Chen, Shibata(2000)]{chen00} 
	Chen, P. F., \& Shibata, K. 2000, \apj, 545, 524
\bibitem[Coppi, Friedland (1971)]{coppi} Coppi, B., \& 
	 Friedland, A. B. 1971, \apj, 169, 379
\bibitem[Cremades, Bothmer(2004)]{cre04}
	Cremades, H. \& Bothmer, V., 2004,  Astron. Astrophys., 422, 307
\bibitem[Feynman, Martin(1995)]{feyn95} 
	Feynman, J. \& Martin, S. F., 1995, \jgr, 100, 3355
\bibitem[Forbes, Acton(1996)]{forb96} 
	Forbes, T. G., \& Acton, L. W., 1996, \apj, 459, 330 
\bibitem[Forbes, Lin(2000)]{forb00}
	Forbes, T. G., \& Lin, J., 2000, J. Atmos. Sol.-Terr. Phys., 62, 1499
\bibitem[Forbes et al.(1989)]{forb89} 
	Forbes, T. G., Malherbe, J. M., \& Priest, E. R. 
	1989, \solphys, 120, 285
%\bibitem[Gosling(1993)]{gos93}
%	Gosling, J. T., 1993, \jgr, 98, 18937
%\bibitem[Glukhov(1997)]{gluk97} 
%        Glukhov, V. S. 1997 \apj, 476, 385
%\bibitem[Hanaoka, et al.(1986)]{hana86} 
%        Hanaoka, Y., Kurokawa, H., \&  Saito, S. 1986, \solphys, 105, 133  
\bibitem[Hanaoka et al.(1994)]{hana94}
        Hanaoka, Y., et al. 1994, \pasj, 46, 205
\bibitem[Harra, Sterling (2001)]{harra01}
	Harra, L. K. \& Sterling, A. C. 2001, \apj, 561, L215
\bibitem[Harrison, Lyons (2000)]{harr00}
	 Harrison, R. A., \& Lyons, M. 2000, Astron. Astrophys., 358, 1097
\bibitem[Hiei et al.(1993)]{hiei93} 
	Hiei, E., Hundhausen, A, J., \& Sime, D. G. 1993, \grl, 20, 2785 
\bibitem[Hirayama(1974)]{hira74}
	Hirayama, T. 1974, \solphys, 34, 323
\bibitem[Hu(1989)]{hu89} 
	Hu, Y. Q. 1989, J. Comput. Phys., 84, 441
%\bibitem[Hudson et al.(1995)]{hud95}
%	Hudson, H., Haisch, B., \& Strong, K. T., 1995, \jgr, 100, 3473
\bibitem[Hudson(1996)]{hud96} Hudson, H. S. 1996, in IAU Colloq. 153
	Magnetodynamic Phenomena in the Solar Atmosphere - 
	Prototyres of Steller Activity, ed. Y. Uchida, T. Kosugi, 
	\& H. S. Hudson (Dordrecht: Kluwer), 89
\bibitem[Hundhausen (1999)]{hund99} 
	Hundhausen, A. 1999, the many faces of the sun : a summary of the 
	results from NASA's Solar Maximum Mission. ed. by Keith T. Strong, 
	Julia L. R. Saba, Bernhard M. Haisch, and Joan T. Schmelz. 
	(New York: Springer), p.143
\bibitem[Illing, Hundhausen (1985)]{ill85}
	Illing, R. M. E. \&  Hundhausen, A. J. 1985, \jgr, 90, 275  
\bibitem[Kahler (1992)]{kahl92}
	Kahler, S. W. 1992, Annu. Rev.  Astron. Astrophys., 30, 113
\bibitem[Lin et al.(2001)]{lin01}
	Lin, J., Forbes, T. G., \& Isenberg, 2001, \jgr, 106, 25053
\bibitem[Lin et al.(2004)]{lin04}
	Lin, J., Raymond, J. C., \& van Ballegooijen, A. A., 2004, 
	\apj, 602, 422
%\bibitem[Low(1994)]{low94} 
%	Low, B. C. 1994, Phys. Plazma, 1, 1684
%\bibitem[MacQueen et al.(1974)]{macq74}
%  MacQueen, R. M., Eddy, J. A., Gosling, J. T., Hildner, E., Munro, R. H., 
%  Newkirk, G. A., Jr., Poland, A. I. \& Ross, C. L., 1974, \apj, 187, L85
\bibitem[McAllister et al.(1996)]{mcal96} 
	McAllister, A. H., Dryer, M., McIntosh, P., Singer, H., \& 
	Weiss, L. 1996, \jgr,   101, 13497 
\bibitem[Magara et al.(1996)]{maga96}
	Magara, T., Mineshige, S., Yokoyama, T., \& Shibata, K., 1996,
	\apj, 466, 1054
%\bibitem[Masuda et al.(1994)]{masu94} 
%	Masuda, S., Kosugi, T., Hara, H., Tsuneta, S., \& Ogawara, Y. 
%	1994, \nat, 371, 495 
\bibitem[Munro et al. (1979)]{munro79}
	Munro, R. H., Gosling, J. T., Hildner, E., MacQueen, R. M., 
	Poland, A. I., \& Ross, C. L., 1979, \solphys, 61, 201 
\bibitem[Nitta et al.(2001)]{nitt01} 
	Nitta, S., Tanuma, S., Shibata, K., \& Maezawa, K. 
	2001, \apj, 550, 1119
%\bibitem[Petschek(1964)]{pet64} 
%	Petschek, H. E. 1964, in Proc. AAS-NASA Simp. on the Physics 
%	of Solar Flares, ed. W. N. Hess(NASA SP-50), 425
\bibitem[Priest (1982)]{prie82}
	Priest, E. R. 1982 Solar Magnetohydrodynamics (Dordrecht: Reidel), 
	chap. 5.4
\bibitem[Raymond, et al.(1976)]{ray76} 
	Raymond, J. C.,  Cox, D. P., \& Smith B. W.  1976, \apj, 204, 290
\bibitem[Sakurai et al.(1992)]{saku92}
	Sakurai, T., Shibata, K., Ichimoto, K., Tsuneta, S.,  \& 
	Acton, L. W. 1992, \pasj, 44, L123 
\bibitem[Sato et al.(1995)]{sato95}
	Sato, T., Hayashi, T., Watanabe, K., Horiushi, R., Tanaka, M., 
	Swairi, N., \&  Kusana, K., 1992, Phis. Fluids B, 4, 450
\bibitem[Shibata et al.(1995)]{shib95} 
	Shibata, K., Masuda, S., Shimojo, M., Hara, H., Yokoyama, T., 
	Tanuma, S., Kosugi, T., \& Ogawara, Y. 1995, \apj, 451, L83
\bibitem[Shibata (1996)]{shib96} 
	Shibata, K., 1996,  Adv. Space Res., 17, 9
\bibitem[Shibata (1999)]{shib99} 
	Shibata, K., 1999, Astrophys. Sp. Sci., 264, 129
%\bibitem[Shibata, Yokoyama(1999)]{shib99} 
%	Shibata, K., \& Yokoyama, T. 1999, \apj, 526, L49
\bibitem[Shiota et al.(2003)]{shiota03}
	Shiota, D. et al. 2003, \pasj, 55, L35
\bibitem[Shiota et al.(2004)]{shiota05}
	Shiota, D. et al. 2004,
   in ASP Conf. Ser. 325, The Solar-B Mission and the Forefront of
   Solar Physics -Proceeding of the Fifth Solar-B Science Meeting-,
   ed. T. Sakurai \& T. Sekii (San Francisco: ASP), 373
\bibitem[Spitzer(1962)]{spit62} 
	Spitzer, L., Jr. 1962 Physics of Fully ionized Gases, 
	Interscience Pub., New York, Chap. 5
\bibitem[Sterling, Hudson (1997)]{ster97}
	Sterling, A. C. \& Hudson, H. S. 1997, \apj, 491, L55
%\bibitem[Tousey (1973)]{tou73}
%	Tousey, R.,1973, Adv. Space Res., 13, 713 
\bibitem[Tsuneta et al.(1991)]{tsu91}
	Tsuneta, S., et al. 1991, \solphys, 136, 37
%\bibitem[Tsuneta et al.(1992a)]{tsu92} 
%	Tsuneta, S., Hara, H., Shimizu, T., Acton, L. W., Strong, K. T., 
%	Hudson, H. S., \& Ogawara, Y. 1992a, \pasj, 44, L63
\bibitem[Tsuneta et al.(1992)]{tsu92} 
	Tsuneta, S., Takahashi, T., Acton, L. W., Bruner, M. E., Harvey, K. L.,
	\& Ogawara, Y. 1992, \pasj, 44, L211
\bibitem[Tsuneta(1996)]{tsu96} 
	Tsuneta, S. 1996, \apj, 456, 840
\bibitem[Ugai(1977)]{ugai77}
	Ugai, M. \& Tsuda, T. 1977, Journal of Pla. Phys., 17, 337 
\bibitem[Webb, Hundhausen (1987)]{webb}
	Webb, D. F. \& Hundhausen, A. J.  \solphys, 1987, 108, 383
\bibitem[Yamamoto et al.(2002)]{yama} 
	Yamamoto, T. T., Shiota, D., Sakajiri, T., Akiyama, S., Isobe, H.,
	\& Shibata, K. 2002, \apj, 579, L45
\bibitem[Yokoyama, Shibata(1997)]{yoko97} 
	Yokoyama, T., \& Shibata, K. 1997, \apj, 474, L61 
\bibitem[Yokoyama,  Shibata(1998)]{yoko98} 
	Yokoyama, T., \& Shibata, K. 1998, \apj, 494, L113 
\bibitem[Yokoyama, Shibata(2001)]{yoko01} 
	Yokoyama, T., \& Shibata, K. 2001, \apj, 549, 1160 
\bibitem[Zarro et al. (1999)]{zarr99}
	Zarro, D. M., Sterling, A. C., Thompson, B. J., Hudson, H. S., \&
	Nitta, N. \apj, 1999, 520, L139
\end{thebibliography}
\end{document}